\documentclass[12pt]{article}
\usepackage{amsmath,amssymb,exscale}
\usepackage{epsfig,psfrag,color,url}
\usepackage{multicol}
\usepackage{multirow}
\input epsf
\newcommand{\m}[1]{\marginpar{{\tiny *}} }
\newcommand{\pslash}{{\not \!p}}
\newcommand{\Dslash}{{\not \!\!D}}
\newcommand{\tr}{{\rm tr}}

\newcommand{\subfigimg}[3][,]{%
  \setbox1=\hbox{\includegraphics[#1]{#3}}
  \leavevmode\rlap{\usebox1}
  \rlap{\hspace*{-2pt}\raisebox{\dimexpr\ht1+0.3\baselineskip}{#2}}
  \phantom{\usebox1}
}

\begin{document}
\topmargin -1.0cm
\oddsidemargin -0.8cm
\evensidemargin -0.8cm

\thispagestyle{empty}

\vspace{40pt}

\begin{center}
\vspace{40pt}

\Large \textbf{A symmetry for $\epsilon_K$}

\end{center}

\vspace{15pt}
\begin{center}
{\large Leandro Da Rold, Iv\'an A. Davidovich} 

\vspace{20pt}

\textit{Centro At\'omico Bariloche, Instituto Balseiro and CONICET}
\\[0.2cm]
\textit{Av.\ Bustillo 9500, 8400, S.\ C.\ de Bariloche, Argentina}

\end{center}

\vspace{20pt}
\begin{center}
\textbf{Abstract}
\end{center}
\vspace{5pt} {\small \noindent
We show a symmetry that, in the context of a composite Higgs with anarchic flavor, can suppress the dominant CP violating contributions to $K-\bar K$ mixing.
Based on previous extensions of SU(3)$_c$, we consider the case in which the composite sector has a global SU(6) symmetry, spontaneously broken to a subgroup containing SU(3)$\times$SU(3). We show that the interactions with the Standard Model can spontaneously break the remaining symmetry to the diagonal subgroup, identified with the group of color interactions, and naturally suppress $\epsilon_K$. We consider this scenario in the context of the Minimal Composite Higgs Model based on SO(5)/SO(4) for the electroweak sector. By working in the framework of 2-site models, we compute the scalar potential, determine the conditions for a successful breaking of the symmetries and calculate the spectrum of lightest states. We find that $\epsilon_K$ can be suppressed and the top mass reproduced for a large region of the parameter space where the symmetries are dynamically broken. Besides other new resonances, the model predicts the presence of a new singlet scalar state, generally lighter than the Higgs, that could have evaded detection at colliders.
}

\vfill
\noindent
{\footnotesize E-mails: daroldl@cab.cnea.gov.ar, ivandavidovich@gmail.com}
\eject

\section{Introduction}
The presence of a new strongly coupled field theory (SCFT) containing a Higgs-like resonance can provide a solution to the electroweak-Planck hierarchy problem as well as a rationale for the flavor hierarchies. One of the most attractive possibilities corresponds to the Higgs being a pseudo Nambu Goldstone Boson (pNGB) arising from the strong dynamics. Besides the Higgs, the SCFT provides new resonances at a scale of few TeV that cut off the momentum integrals in the Higgs potential. Generically, the presence of these states induce corrections to the electroweak (EW) observables, that are in conflict with EW precision tests. Many of the constraints can be alleviated by the use of suitable symmetries~\cite{Agashe:2003zs,Agashe:2006at}.

A key ingredient to obtain flavor hierarchies in composite Higgs models is partial compositeness~\cite{Kaplan:1991dc,Contino:2004vy}. In partial compositeness hierarchies in the fermion masses can be easily obtained by assuming that the degree of compositeness of the fermions are hierarchical, without hierarchies or relations in the flavor structure of the SCFT. This assumption leads to non-universal couplings between the SM fermions and the resonances of the SCFT, such that virtual exchange of resonances induces flavor violating effects. Despite this violation, partial compositeness contains a built-in GIM-like mechanism suppressing flavor changing neutral currents (FCNC)~\cite{Gherghetta:2000qt,Agashe:2004cp}. This setup is known as the anarchic approach to the flavor problem.

Although the GIM-like mechanism successfully suppresses most of the contributions to FCNC, the very stringent constraints from Kaon physics, in particular $\epsilon_K$, push the scale of the composite resonances to $m_{cp}\sim 10\div 30$ TeV, worsening the fine-tuning problem in the EW sector~\cite{Csaki:2008zd,Redi:2011zi,Panico:2015jxa}. Some alternatives to avoid these constraints have been considered in the literature, such as the presence of flavor symmetries~\cite{Csaki:2008zd,Redi:2011zi} and hierarchies in the SCFT~\cite{Delaunay:2010dw}, as well as the presence of different dynamical scales~\cite{Panico:2016ull}. These scenarios relax the constraints allowing $m_{cp}$ to be lowered to a few TeV. In the context of anarchic flavor, the authors of Ref.~\cite{Bauer:2011ah} have shown that enhancing the global symmetry of the SCFT from SU(3)$_c$ to SU(3)$_L\times$SU(3)$_R$, where the indices can be associated with the chiralities of the quarks, the scale can be lowered to $m_{cp}\sim 2$ TeV. In that scenario, masses for the SM quarks require an extension of the scalar sector of the theory, with new scalar fields charged under the extended symmetry. The masses of these states are expected to be of the same order as $m_{cp}$, introducing a sizable breaking of the enlarged symmetry and spoiling the mechanism protecting $\epsilon_K$, unless a fine-tuning of several orders of magnitude is introduced~\cite{DaRold:2012sz}.

In the present paper we will elaborate on the possibility of protecting $\epsilon_K$ from large corrections by considering an extension of SU(3)$_c$, within the context of an anarchic SCFT. We will work along the lines of Ref.~\cite{Bauer:2011ah}, but making the new scalar a pNGB, based on a global symmetry SU(6) spontaneously broken by the strong dynamics to a subgroup containing SU(3)$_L\times$SU(3)$_R$. Besides this extension of the color symmetry group, we add the usual SO(5)$\times$U(1)$_X$ global symmetry that allows us to describe also the Higgs field as a pNGB. The interactions between the SCFT and the SM gauge and fermion fields generate a potential for both scalars at one-loop. The potential can trigger electroweak symmetry breaking (EWSB) and it can also spontaneously break SU(3)$_L\times$SU(3)$_R$ to the diagonal subgroup that is aligned with SU(3)$_c$. The size of the spontaneous breaking can be described in terms of the parameters $\epsilon_5$ and $\epsilon_6$, that take values $0\leq\epsilon_{5,6}\leq 1$, and measure the misalignment of the vacuum in the direction of SO(5) and SU(6), respectively. As in the usual Minimal Composite Higgs Model (MCHM): $v_{SM}=\epsilon_5 f_5$, with $f_5$ the decay constant of the pNGB Higgs~\cite{Agashe:2004rs}. After spontaneous breaking of both symmetries fermion masses are successfully generated, leading to a realistic model. Quark masses, requiring breaking of both symmetries, are proportional to $\epsilon_5\epsilon_6$. The dominant effect in $\epsilon_K$ is given by a contribution to the Wilson coefficient $C_4$ that, compared with the case without extended symmetry, is suppressed by $\epsilon_6^2(1-\epsilon_6^2)$. In general, factors of $\epsilon_6\sqrt{1-\epsilon_6^2}$ are expected in operators with $LR$ chiral structure, as for example magnetic dipole moments.

The values of $\epsilon_5$ and $\epsilon_6$ depend on the parameters of the theory, as well as on the representations of the fermions under SO(5) and SU(6). The representations of SU(6) are fixed to cancel the dominant contributions to $\epsilon_K$, according to~\cite{Bauer:2011ah}. We discuss the embedding into SO(5) representations, and show that a realistic case is obtained for $q_L$ and $u_R$ embedded in the representations 14 and 1 of SO(5) respectively. With this choice the potential can easily trigger spontaneous breaking of both symmetries, generating masses for the SM fields. 

We use a very simple description of the above dynamics by considering a 2-site model; that is, a four dimensional theory containing the first level of resonances of the SCFT~\cite{Contino:2006nn,Foadi:2010bu,Panico:2011pw,DeCurtis:2011yx}. Within this framework we are able to calculate the one-loop Coleman-Weinberg potential that is dominated by the top and gauge contributions. We compute $\epsilon_5$ and $\epsilon_6$, as well as the masses and mixing angle of the light scalars: the Higgs and the new neutral state. We find that, taking $f\simeq 1$ TeV, the Higgs is usually somewhat heavier than 125 GeV, as expected in models with fermions in the 14. The other scalar is usually lighter, with a mass of order $50\div 100$ GeV due to a suppressed quartic coupling, whose size is determined by the embedding of the quarks into the SU(6) symmetry. 

The paper is organized as follows: In sec.~\ref{sec-extended-sym} we describe the symmetries of the theory assuming that the Higgs and the new scalar are pNGBs of a SCFT; we show the effective theory that is obtained after the heavy resonances of the SCFT are integrated out, codifying their effects in terms of a set of form factors. In sec.~\ref{sec-potential} we compute the one-loop potential by using the form factors of the effective theory, we analyse the conditions for spontaneous symmetry breaking and the spectrum of light scalar states; we also discuss the dependence of the potential with the SO(5) embedding of the fermions. In sec.~\ref{sec-2-site} we describe a 2-site model that provides a calculable framework for the above dynamics and we compute the contributions to some flavor observables. In sec.~\ref{sec-num-results} we show numerical results for several interesting physical quantities, namely the spectrum of light states and the regions of the parameter space with spontaneous breaking of the symmetries. In sec.~\ref{sec-discussions} we discuss some important issues, such as the neutron dipole moments, the degree of tuning of the model and the phenomenology of the new states at the LHC.

\section{A symmetry for $\epsilon_K$}\label{sec-extended-sym}
We consider a new strongly interacting sector beyond the SM. It has a global symmetry SO(5)$\times$U(1)$_X\times$SU(6).~\footnote{As is well known, the extra U(1)$_X$ factor is required to obtain the proper normalization of hypercharge in the fermionic sector.} The interactions produce bound states with masses of order TeV and spontaneously break the global symmetry to SO(4)$\times$U(1)$_X\times$SU(3)$_L\times$SU(3)$_R\times$U(1). A set of NGBs $\Pi_5$ and $\Pi_6$ transforming non-linearly under the global SO(5) and SU(6), respectively, emerge from this breaking. The field $\Pi_5$ has the proper quantum numbers to be identified with the Higgs multiplet. The field $\Pi_6$ results in a new scalar field, with some components charged under the color interactions of the SM. The SM fermions and gauge bosons are elementary fields, and can be thought of as external sources probing the strong dynamics. Depending on the orientation of the vacuum compared to the SM gauge symmetry SU(2)$_L\times$U(1)$_Y\times$SU(3)$_c$, the EW and color symmetries can be broken or not. This direction is determined dynamically by the potential generated at loop level from the interactions between the SM fields and the strongly interacting sector that explicitly violate the global symmetry. The presence of this potential turns $\Pi_5$ and $\Pi_6$ into pNGBs.

The interactions between the SM gauge fields and the strongly interacting sector are introduced by the gauging of a subgroup of the global symmetry; in this way the elementary gauge fields are coupled to a subset of the conserved currents of the composite sector. The elementary fermions are coupled linearly with the strongly interacting sector, realizing partial compositeness.

To shorten the notation we define G$_6\equiv$SU(6) and G$_5\equiv$SO(5)$\times$U(1)$_X$, and for the subgroups H$_6\equiv$SU(3)$_L\times$SU(3)$_R\times$U(1) and H$_5\equiv$SO(4)$\times$U(1)$_X$. 
A possible orientation of the vacuum in the direction of G$_6$ can be described by the following matrix:
\begin{equation}\label{eq-Phi0}
\Omega_0=\left(\begin{array}{cc} I_3 & 0\\ 0 & -I_3 \end{array}\right)
\end{equation}
where $I_3$ is the identity matrix in three dimensions. Under G$_6$ this vacuum transforms as: $g_6\Omega_0g_6^\dagger$, with $g_6$ an element of G$_6$ in the fundamental representation. This vacuum preserves an H$_6$ subgroup. We identify SU(3)$_c$ with the diagonal subgroup SU(3)$_V$ contained in H$_6$. Therefore, expanding around this vacuum one can classify the fields according to their representation properties under the SM group of strong interactions. 

The NGBs $\Pi_6$ are given by a local element of G$_6$ of the form:
\begin{equation}
U_6=e^{i\sqrt{2}\Pi_6/f_6}\ ,\qquad \Pi_6=\Pi_6^{\hat b}T_6^{\hat b} \ ,
\end{equation}
with $T_6^{\hat b}$ the broken generators of G$_6$. We will use the subindex 6 for elements related with this group, to distinguish them from the elements associated to G$_5$. Only elements of the form $U_6$ transform the vacuum $\Omega_0$. $U_6$ transforms non-linearly under G$_6$, as: $g_6\ U_6\ h_6^\dagger(g_6,\Pi_6)$, with $h_6\in$H$_6$. $\Pi_6$ transforms linearly under H$_6$, as a complex $(\bar 3,3)_{1/\sqrt{3}}$. We find it useful to define the matrix $\Phi_6=\Pi_6|_{\rm fund}$, where the subindex indicates that the generators $T_6^{\hat b}$ are in the fundamental representation. For the suitable basis defined in Ap.~\ref{Ap-SU(6)}, $\Phi_6$ can be written as:
\begin{equation}
\Phi_6=\left(\begin{array}{cc} 0 & \Phi_{(\bar 3,3)}\\ \Phi_{(\bar 3,3)}^\dagger & 0 \end{array}\right)
\end{equation}
with $\Phi_{(\bar 3,3)}$ a $3\times 3$ complex matrix with eighteen real degrees of freedom, describing the NGB fields.

A representative orientation of the vacuum in the direction of G$_5$ can be parametrized by the vector~\cite{Agashe:2004rs}
\begin{equation}\label{eq-sigma0}
\Sigma_0=(0,0,0,0,1) \ .
\end{equation}
The vacuum $\Sigma_0$ transforms as: $g_5\Sigma_0$, with $g_5$ an element of SO(5) in the fundamental representation, $\Sigma_0$ is invariant under an H$_5$ subgroup. We identify the generators of the EW symmetry of the SM with a subset of the generators of H$_5$. Expanding around this vacuum all the fields can be classified according to their representations under the EW gauge symmetry. 

The NGB of G$_5/$H$_5$ are given by:
\begin{equation}
U_5=e^{i\sqrt{2}\Pi_5/f_5}\ ,\qquad \Pi_5=\Pi_5^{\hat a}T_5^{\hat a}
\end{equation}
with $T_5^{\hat a}$ the broken generators of G$_5$. In this case we use the subindex 5 for elements related with G$_5$. Only elements of the form $U_5$ transform the vacuum. $U_5$ transforms non-linearly under G$_5$, as: $g_5\ U_5\ h_5^\dagger(g_5,\Pi_5)$, with $h_5\in$H$_5$, whereas $\Pi_5$ transforms with the fundamental representation $(2,2)_0$ of H$_5$. Similar to the case of $\Phi_6$, we define the matrix $\Phi_5=\Pi_5|_{\rm fund}$, with the generators $T_5^{\hat a}$ in the fundamental representation.

As we will show in sec.~\ref{sec-potential}, virtual exchange of elementary fields can misalign the vacuum at loop level, breaking H$_5$ to K$_5\equiv$SO(3) and H$_6$ to K$_6\equiv$SU(3)$_V$. Under K$_5$ the NGB $\Pi_5$ decomposes as: $(2,2)\sim 1\oplus 3$, whereas under K$_6$ the NGB $\Pi_6$ decomposes as: $(\bar 3,3)\sim 1\oplus 8$, with the singlet and octet being complex fields. The singlets of $\Pi_5$ and $\Pi_6$ under K$_5$ and K$_6$ will be called $h_5$ and $h_6 e^{i\theta_6}$, respectively, and we define:
\begin{align}\label{def-s-eps}
s_5 &= \sin(h_5/f_5), \qquad &s_6 = \sin(h_6/\sqrt{6}f_6), \\
\epsilon_5 &=\sin(v_5/f_5), \qquad &\epsilon_6 =\sin(v_6/\sqrt{6}f_6),
\end{align}
with $v_5=\langle h_5\rangle$ and $v_6=\langle h_6\rangle$ the vacuum expectation values of the singlets. By using these definitions the vacuum can be characterized by the following objects:
\begin{align}\label{eq-vacuum}
\Sigma_v=(0,0,0,\epsilon_5,\sqrt{1-\epsilon_5^2}) \ , \qquad
\Omega_v=\left(\begin{array}{cc} (1-2\epsilon_6^2) I_3 & i2\epsilon_6\sqrt{1-\epsilon_6^2}\ I_3\\ i2\epsilon_6\sqrt{1-\epsilon_6^2}\ I_3 & -(1-2\epsilon_6^2) I_3 \end{array}\right) \ .
\end{align}
Thus $\epsilon_5$ and $\epsilon_6$ measure the misalignment of the vacuum, with $\epsilon_5=\epsilon_6=0$ in the case of neither EWSB nor H$_6$ breaking, and $\epsilon_5=\epsilon_6=1$ in the case of maximal symmetry breaking of both groups.

\subsection{Effective theory}\label{sec-eff-th}
Integrating out the heavy degrees of freedom of the strongly interacting sector one can obtain an effective theory for the elementary degrees of freedom and the NGBs. We find it useful to add spurious degrees of freedom to the elementary sector to extend its symmetry to G$_5\times$G$_6$. Therefore one can write an effective Lagrangian that is formally invariant under this extended symmetry. In this procedure one has to choose the representations of the elementary fermions. We choose the fundamental representation of G$_6$ for the quarks, which under H$_6$ decomposes as: $6\sim (3,1)_{1/2\sqrt{3}}\oplus(1,3)_{-1/2\sqrt{3}}$. We choose (3,1) for $q_L$ and (1,3) for $q_R$ to suppress the main contribution to $\epsilon_K$~\cite{Bauer:2011ah}. The representations of G$_5$ are not fixed, the only constraint being that they have to contain the proper doublets and singlets of SU(2)$_L$ to match with the SM fields. As is well known, the choice of the representations of G$_5$ has important consequences for the phenomenology; we will discuss their impact in the present model in sec.~\ref{sec-potential}.

As usual, due to the non-linear transformation properties of the NGB, a G$_5\times$G$_6$ invariant Lagrangian can be obtained by dressing the elementary fields with the NGB matrices U$_5$ and U$_6$, and building with them a Lagrangian that superficially looks invariant under H$_5\times$H$_6$. 

Calling $R$ a representation of G$_5$, $R$ decomposes under H$_5$ as $R\sim\oplus_j r_j$. Similarly, calling $S$ a representation of G$_6$, $S$ decomposes under H$_6$ as $S\sim\oplus_k s_k$. Therefore an elementary fermion $\psi$ in the representation $R$ of G$_5$ and $S$ of G$_6$, decomposes under H$_5\times$H$_6$ as: $\psi\sim\oplus_{r,s}\psi_{rs}$. Dressing $\psi$ with the NGB matrices usually leads to fields that transform with reducible representations, therefore the building blocks for the effective theory are the projections of the dressed fields into the irreducible representations of the unbroken subgroup: $\tilde\psi_{rs}=(U_5^\dagger U_6^\dagger \psi)_{rs}$, where the generators contained in $U_5$ are in the representation $R$ of G$_5$ and the generators contained in $U_6$ are in the fundamental representation of G$_6$, as discussed previously. G$_5\times$G$_6$ invariant operators can be obtained by considering products of factors $\tilde\psi_{rs}$, and selecting from these products the invariants of H$_5\times$H$_6$. Using these objects, at the quadratic level in the elementary fields, the effective Lagrangian for the quarks is:
\begin{align}\label{eq-leff-f}
{\cal L}_{\rm eff}\supset&\sum_{f=q,u,d}Z_f\bar\psi_f \pslash\psi_f+
\sum_{f=q,u,d}\sum_{r,s}\overline{(U_5^\dagger U_6^\dagger\psi_f)}_{rs}\ \pslash\ \Pi^f_{rs}(p^2)\ (U_5^\dagger U_6^\dagger\psi_f)_{rs} \nonumber\\
&+ \sum_{f=u,d}\sum_{r,s}\overline{(U_5^\dagger U_6^\dagger \psi_q)}_{rs}\ M^f_{rs}(p^2)\ (U_5^\dagger U_6^\dagger \psi_f)_{rs} + {\rm h.c.} \ ,
\end{align}
where $Z_f$ are the factors of the elementary kinetic terms. $\Pi^f_{rs}(p^2)$ and $M^f_{rs}(p^2)$ are momentum dependent form factors that contain the information on the composite degrees of freedom; they are independent of the NGB fields. The components of the different factors of the dressed fermions $\tilde\psi_{rs}$ in Eq.~(\ref{eq-leff-f}) are contracted in the usual way to obtain H$_5\times$H$_6$ invariants. Generation indices are understood in Eq.~(\ref{eq-leff-f}), and in equations below.

For the gauge sector, it is useful to notice that the adjoint representation of SO(5) decomposes under SO(4) as: $10\sim\oplus_i t_i = (3,1)\oplus(1,3)\oplus(2,2)$, whereas the adjoint of G$_6$ decomposes under H$_6$ as: $35\sim \oplus_j u _j = (8,1)_0\oplus(1,8)_0\oplus(\bar 3,3)_{1/\sqrt{3}}\oplus(1,1)_0$. Calling $x_\mu$, $a_\mu$ and $g_\mu$ the fields of U(1)$_x$, SO(5) and G$_6$, respectively, the quadratic effective Lagrangian for the elementary gauge fields is:
\begin{align}\label{eq-leff-g}
{\cal L}_{\rm eff}\supset \frac{1}{2}P^{\mu\nu}\left[\Pi^X(p^2)x_\mu x_\nu + \sum_t\Pi^A_t(p^2)(U_5^\dagger a_\mu)_t\ (U_5^\dagger a_\nu)_t + \sum_u\Pi^G_u(p^2)(U_6^\dagger g_\mu)_u\ (U_6^\dagger g_\nu)_u \right]\ ,
\end{align}
with $P_{\mu\nu}=\eta_{\mu\nu}-p_\mu p_\nu/p^2$, $U_5$ and $U_6$ in the adjoint representation, and $t$ and $u$ running over the decompositions of each adjoint under the subgroups as presented in the previous paragraph. There are also elementary kinetic terms for the gauge fields that were not written in Eq.~(\ref{eq-leff-g}). The form factors $\Pi(p^2)$ encode the effects of the strongly coupled sector and are independent of the NGB fields. 

There are also kinetic terms for the NGB fields that can be constructed, as usual, by making use of the Maurer-Cartan form $d_\mu$ defined from: $iU^\dagger D_\mu U=e_\mu^aT^a+d_{\mu\hat a}T^{\hat a}\equiv e_\mu+d_\mu$. There is a $d_\mu^5$ for the NGBs $\Pi_5$ and a $d_\mu^6$ for the NGBs $\Pi_6$. The NGBs kinetic terms are given by:
\begin{equation}\label{eq-leff-Pi}
{\cal L}_{\rm eff}\supset \frac{f_5^2}{4}d^{\hat a}_{5,\mu} d_{5}^{\hat a,\mu} + \frac{f_6^2}{4}d^{\hat b}_{6,\mu} d_{6}^{\hat b,\mu}  \ .
\end{equation}

Keeping only the dynamical fermions: $q_L=(u_L,d_L)$ contained in $\psi_q$, as well as $u_R$ and $d_R$ contained in $\psi_u$ and $\psi_d$, Eq.~(\ref{eq-leff-f}) leads to:
\begin{align}\label{eq-leff-f-dyn}
{\cal L}_{\rm eff}\supset&\, \bar q_L \pslash\ \Pi_q(\Pi_5,\Pi_6,p^2) q_L+\bar u_R\pslash\ \Pi_u(\Pi_5,\Pi_6,p^2) u_R+\bar d_R\pslash\ \Pi_d(\Pi_5,\Pi_6,p^2) d_R \nonumber\\ & +\bar q_LM_u(\Pi_5,\Pi_6,p^2)u_R+\bar q_LM_d(\Pi_5,\Pi_6,p^2)d_R + {\rm h.c.} \ .
\end{align}
Since $\Pi(\Pi_5,\Pi_6,p^2)$ and $M(\Pi_5,\Pi_6,p^2)$ depend on momentum and on the NGBs, they have non-trivial indices under SU(3)$_c\times$SU(2)$_L$. Using the form factors defined in Eq.~(\ref{eq-leff-f}), they can be written as:
\begin{align}\label{eq-pi1}
&\Pi_f(\Pi_5,\Pi_6,p^2)=Z_f+\sum_{r,s}\Pi^f_{rs}(p^2)F^f_{rs}(\Pi_5,\Pi_6) \ , & f=q,u,d\ ,\\
&M_f(\Pi_5,\Pi_6,p^2)=\sum_{r,s}M^f_{rs}(p^2)G^f_{rs}(\Pi_5,\Pi_6) \ ,  & f=u,d \ ,\label{eq-pi2}
\end{align}
where all the momentum dependence is codified in the form factors defined in Eq.~(\ref{eq-leff-f}), and the NGB dependence is contained in the functions $F^f_{rs}$ and $G^f_{rs}$. These functions can be obtained from the invariants built with the dressed quarks:
\begin{align}\label{eq-funPi}
& \left[F^q_{rs}(\Pi_5,\Pi_6)\right]_{\alpha\beta ij}=\partial_{\bar q_L^{\alpha i}}\partial_{q_L^{\beta j}}\left[\overline{(U_5^\dagger U_6^\dagger\psi_q)}_{rs}\ (U_5^\dagger U_6^\dagger\psi_q)_{rs}\right] \ , \nonumber \\
& \left[F^f_{rs}(\Pi_5,\Pi_6)\right]_{\alpha\beta}=\partial_{\bar f_R^{\alpha}}\partial_{f_R^{\beta}}\left[\overline{(U_5^\dagger U_6^\dagger\psi_f)}_{rs}\ (U_5^\dagger U_6^\dagger\psi_f)_{rs}\right] \ , \qquad f=u,d \ , \nonumber \\
& \left[G^f_{rs}(\Pi_5,\Pi_6)\right]_{\alpha ij}=\partial_{\bar q_L^{\alpha i}}\partial_{f_R^{\beta}}\left[\overline{(U_5^\dagger U_6^\dagger\psi_q)}_{rs}\ (U_5^\dagger U_6^\dagger\psi_f)_{rs}\right] \ , \qquad f=u,d \ ,
\end{align}
where $\alpha,\beta=1,2,3$ are color indices, and $i,j=1,2$ are SU(2)$_L$ indices. It is straightforward to show that these functions factorize in one factor containing the $\Pi_5$-dependence and another factor containing the $\Pi_6$-dependence: $F^f_{rs}(\Pi_5,\Pi_6)=F^f_{r}(\Pi_5)F^f_{s}(\Pi_6)$ and similarly for $G$.

The dynamical gauge fields of QCD are associated to the generators of the diagonal subgroup $T_6^{V,a}\equiv T_6^{L,a}+T_6^{R,a}$, for $a=1,\dots 8$. In this case Eq.~(\ref{eq-leff-g}) leads to:
\begin{equation}\label{eq-leff-g-dyn}
{\cal L}_{\rm eff}\supset \frac{1}{2}P^{\mu\nu}g^{V,a}_\mu\Pi_G^{ab}(\Pi_6,p^2)g^{V,b}_\nu\ ,
\end{equation}
where $\Pi_G(\Pi_6,p^2)$ is given by:
\begin{equation}\label{eq-pig1}
\Pi_G(\Pi_6,p^2)=Z_6+\sum_{s}\Pi^G_s(p^2)F^G_s(\Pi_6) \ ,
\end{equation}
with $Z_6=g_{6,0}^{-2}$ being the coefficient of the elementary kinetic term, and
\begin{align}\label{eq-funPig}
\left[F^G_s(\Pi_6)\right]_{ab}=\partial_{g^{V,a}}\partial_{g^{V,b}}\left[(U_6^\dagger g^V)_s\ (U_6^\dagger g^V)_s\right] \ .
\end{align}

A similar description can be done for the EW sector~\cite{Agashe:2004rs,Andres:2015oqa}.

In order to obtain the effective Lagrangian at quadratic level, it is useful to evaluate the scalar fields to their vev in Eq.(\ref{eq-leff-f-dyn}). In this case the functions $F^f_{rs}$ of Eq.~(\ref{eq-pi1}) are proportional to the identity in color space~\footnote{$\Pi_q$ splits into four $3\times 3$ blocks, with those in the diagonal proportional to the identity and the others vanishing.}, thus they are color independent, as are the form factors. The functions $G^f_{rs}$ of Eq.~(\ref{eq-pi2}) split into two $3\times 3$ blocks, one proportional to the identity and the other vanishing. Moreover, the NGB matrices $U_5$ and U$_6$ can be resummed, leading to matrices that can be expressed as functions of $\epsilon_5$ and $\epsilon_6$. Taking into account the representations of the quarks under G$_6$ and H$_6$, and for generic representations of G$_5$ and H$_5$, the form factors can be written as:
\begin{align}
&\Pi_{f_L}(p^2)=Z_q+\sum_{r}\Pi^q_{r31}(p^2)\ F^{f_L}_r(\epsilon_5)+\epsilon_6^2\sum_{r}[\Pi^q_{r13}(p^2)-\Pi^q_{r31}(p^2)]\ F^{f_L}_r(\epsilon_5) \ , \nonumber\\
&\Pi_{f_R}(p^2)=Z_f+\sum_{r}\Pi^f_{r13}(p^2)\ F^{f_R}_r(\epsilon_5)+\epsilon_6^2\sum_{r}[\Pi^f_{r31}(p^2)-\Pi^f_{r13}(p^2)]\ F^{f_R}_r(\epsilon_5) \ , \nonumber\\
&\Pi_{f_Lf_R}(p^2)=-i\epsilon_6\sqrt{1-\epsilon_6^2}\sum_{r}[M^f_{r31}(p^2)-M^f_{r13}(p^2)]\ G^{f_Lf_R}_r(\epsilon_5) \ , \qquad f=u,d\ , \label{eq-corr-f-phys}
\end{align}
where 31 and 13 label the H$_6$ representations $(3,1)$ and $(1,3)$, respectively, omitting the parenthesis to simplify the notation.
The functions $F_r(s_5)$ can be obtained from Eqs.~(\ref{eq-pi1}) and~(\ref{eq-pi2}), and have been computed for several representations in the literature, see for example Refs.~\cite{Contino:2006qr,Montull:2013mla,Carena:2014ria,Andres:2015oqa} and Ap. \ref{invariants}.

As usual, the spectrum of fermions that mix with the SM ones corresponds to the zeroes of the following function:
\begin{equation}\label{eq-fermion-spectrum}
{\rm Zeroes}[\Pi_{f_L}(p^2)\Pi_{f_R}(p^2)-|\Pi_{f_Lf_R}(p^2)|^2] \ .
\end{equation}
The zero of each function that is closest to the origin gives the mass of the corresponding SM fermion.

\subsection{Partial compositeness and flavor}
Besides the group theoretical structure of the correlators, another useful piece of information arises from the hypothesis of partial compositeness~\cite{Kaplan:1991dc,Contino:2004vy}. We define $\epsilon_f$ as the degree of compositeness of the fermions, with $\epsilon_f\ll 1$ for mostly elementary fermions and $\epsilon_f\sim 1$ for partially composite fermions~\cite{Contino:2006nn}. Thus, to leading order in $\epsilon_f$: $\Pi^f\propto \epsilon_f^2$ and $M^f\propto\epsilon_q\epsilon_f$. 

There are two sources of flavor structure: the structure of the SCFT and the structure of partial compositeness contained in $\epsilon_f$. Flavor anarchy corresponds to the assumption of anarchic flavor structure of the SCFT, meaning that there are no hierarchies in the couplings and masses of this sector. As is well known, in this case the hierarchies between the quark masses can be obtained by hierarchical degrees of compositeness $\epsilon_f$. In addition, this mechanism can also lead to a realistic $V_{CKM}$. The hierarchy in $\epsilon_f$ can be naturally obtained in five dimensional models~\cite{Grossman:1999ra,Gherghetta:2000qt,Contino:2004vy} and in cases where the SCFT has an approximate conformal symmetry at high energies~\cite{Luty:2004ye,Rattazzi:2008pe}. We will show more details in sec.~\ref{sec-2-site}.

A useful approximate expression for the masses of the SM fermions can be derived from the effective Lagrangian~\cite{Agashe:2004rs}. By canonically normalizing the kinetic term of the fermions and evaluating the scalars to their vev, we obtain:
\begin{equation}\label{eq-mu}
m_f\simeq\frac{|\Pi_{f_Lf_R}(0)|}{\left[\Pi_{f_L}(0)\Pi_{f_R}(0)-2|\Pi_{f_Lf_R}(0)|\left.\frac{d |\Pi_{f_Lf_R}(p^2)|}{d p^2}\right|_{p^2=0}\right]^{1/2}} \ ,
\end{equation}
where $\Pi_{f_L}$, $\Pi_{f_R}$ and $\Pi_{f_Lf_R}$ have been defined in Eq.~(\ref{eq-corr-f-phys}). Notice that, as expected, besides EWSB fermion masses require breaking of H$_6$. Also, for maximal breaking of H$_6$: $\epsilon_6=1$, the fermion masses vanish. Partial compositeness shows us that, to leading order in the mixing, the masses are of order
\begin{equation}\label{eq-mu-2}
m_f\sim \epsilon_6\sqrt{1-\epsilon_6^2}\, G^{f_Lf_R}_{r}(\epsilon_5)\, \epsilon_q\, m_r\, \epsilon_f \ ,
\end{equation}
with $m_r$ the mass scale of the fermionic resonances of the SCFT. All the group structure is contained in the first three factors, whereas the flavor structure is contained in $\epsilon_q\epsilon_f m_r$. Comparing with the usual result in composite Higgs models, there is an extra factor $\epsilon_6\sqrt{1-\epsilon_6^2}$ that arises from the extended symmetry G$_6$.

The Yukawa couplings with the scalars $h_5$ and $h_6$ can be computed by taking the derivative of $m_f$, as approximated in Eq.~(\ref{eq-mu}), with respect to $v_5$ and $v_6$, respectively:
\begin{equation}\label{eq-def-yukawa}
y_{f,5}\simeq\frac{\partial m_f}{\partial v_5} \ ,\qquad y_{f,6}\simeq\frac{\partial m_f}{\partial v_6} \ .
\end{equation}
The dependence of the Yukawa couplings on the parameters of the theory is encoded in the form factors, but the dependence on the vev's of the scalar fields is simple, it is encoded in $\epsilon_6$ and in the functions $F_r(\epsilon_5)$ and $G_r(\epsilon_5)$ of Eq.~(\ref{eq-corr-f-phys}). These functions are straightforward to compute once the embedding of the fermions on G$_5$ is chosen, see Ap.~\ref{invariants}. A good estimate of the size of the Yukawa couplings can be obtained by expanding Eq.~(\ref{eq-def-yukawa}) to leading order in powers of $1/Z_f$, similar to an expansion in powers of the fermion mixing $\epsilon_f$:
\begin{align}
&y_{f,5}\simeq m_f\ \frac{\sqrt{1-\epsilon_5^2}}{f_5}\ \frac{d}{d\epsilon_5}\log\sum_{r}[M^f_{r31}(0)-M^f_{r13}(0)]\  G_r^{f_Lf_R}(\epsilon_5), \label{eq-yf5-aprox}
\\
&y_{f,6}\simeq m_f\frac{1}{\sqrt{6}f_6}\frac{1-2\epsilon_6^2}{\epsilon_6\sqrt{1-\epsilon_6^2}}\ .\label{eq-yf6-aprox}
\end{align}
A non-vanishing $y_{f,5}$ requires $0<\epsilon_6<1$. In the cases with just one non-trivial invariant function $G_r$, the dependence of the derivative on the form factors cancels out, and $y_{f,5}$ depends on the microscopic parameters of the theory only through $\epsilon_5$ and the combination $m_f/f_5$. The predictions for $y_{f,5}$ in the case of representations with custodial symmetry of $Zb_L\bar b_L$ have been reported in Refs.~\cite{Montull:2013mla,Carena:2014ria}. On the other hand, a non-vanishing $y_{f,6}$ requires $\epsilon_5>0$. Also, $y_{f,6}$ adopts a very simple form in this expansion, since all the dependence on the microscopic parameters of the theory has been absorbed in $m_f/f_6$, and the dependence on $\epsilon_6$ is shown explicitly. This dependence is determined by the embedding of the quarks in the extended symmetry G$_6$.

We are also interested in contributions to the Wilson coefficients of four-fermion operators, in particular to the operator ${\cal O}_4=(\bar \psi_q\psi_d)(\bar\psi_d\psi_q)$. 
However, instead of keeping track of the full NGB dependence, we are interested only in the coefficient evaluated in the vacuum of Eq.~(\ref{eq-vacuum}), and keeping the dynamical fermions only. We obtain: 
\begin{equation}
{\cal L}_{\rm eff}\supset  \epsilon_6^2 (1-\epsilon_6^2) \, \Pi_{f_Lf_R}^{(4)}(p^2) (\bar f_L f_R)(\bar f_R f_L) \ ,
\end{equation}
where $\Pi_{f_Lf_R}^{(4)}(p^2)$ is a form factor independent of $\epsilon_5$ and $\epsilon_6$, and where flavor indices are understood. The Wilson coefficient $C_4$ is
\begin{equation}\label{eq-C4eff}
C_4=\epsilon_6^2 (1 - \epsilon_6^2) \, \Pi_{f_Lf_R}^{(4)}(0) \ .
\end{equation}
Partial compositeness implies that $\Pi^{(4)}_{f_Lf_R}$, and consequently also $C_4$, is proportional to $(\epsilon_{f_L}\epsilon_{f_R})^2$.
Eq.~(\ref{eq-C4eff}) must be compared with the usual result in composite Higgs models, without the extended symmetry G$_6$, where the factor $\epsilon_6^2 (1 - \epsilon_6^2)$ is absent. In that case the bounds on the scale of the masses of the composite resonances arising from flavor physics are $m_{cp}\gtrsim 10\div 30$TeV~\cite{Panico:2015jxa}. In the present case, the extended symmetry of the SCFT gives an extra suppression factor to the contribution to $C_4$, that can alleviate this bound. Up to corrections of numerical factors of ${\cal O}(1)$, that depend on the specific realization of the SCFT, for $\epsilon_6\sqrt{1 - \epsilon_6^2}\simeq 0.1\div 0.3$, the scale can be as low as $m_{cp}\sim 3$~TeV. Moreover, as we will show in sec.~\ref{sec-2-site}, in 2-site models there is an extra factor 1/2 that relaxes $\epsilon_6\sqrt{1 - \epsilon_6^2}\lesssim 0.2\div 0.5$ for $m_{cp}\sim 3$~TeV. Notice that not only a small $\epsilon_6$ can suppress the contribution to $C_4$, also for large $\epsilon_6$, near one, there can be a supression.

The same factor involved in $C_4$: $\epsilon_6  \sqrt{1 - \epsilon_6^2}$, suppresses the fermion masses, see Eq.~(\ref{eq-mu-2}). Thus there is a tension between the large top mass and a small $C_4$. We will elaborate more on this topic in sec.~\ref{sec-2-site}.

\section{Radiative Potential}\label{sec-potential}
The interactions between the fermions of the SM and the strongly interacting sector explicitly break G$_5\times$G$_6$ to the subgroup SU(2)$_L\times$U(1)$_Y\times$H$_6$, generating a potential for the NGB fields at the one-loop level. This potential can misalign the vacuum and induce EW and H$_6$ spontaneous breaking. The interactions between the color gauge fields of the SM and the SCFT explicitly break G$_6$ to SU(3)$_V$, whereas the interactions with the EW gauge fields break G$_5$ to SU(2)$_L\times$U(1)$_Y$. In general the interactions with the gauge fields do not induce misalignment of the vacuum. Below we discuss the contributions of the fermions, as well as the contributions of the gauge sector with color. The contributions of the EW gauge fields have been discussed extensively; we refer the reader to the original articles on SO(5)/SO(4)~\cite{Agashe:2004rs}.

Let us start with the fermion contribution. We will consider only the effect of $q_L$ and $u_R$ of the third generation, since, having the largest mixing, they give the dominant contribution. We obtain:
\begin{equation}\label{vf}
V_f(\Pi_5,\Pi_6)=-2\int \frac{d^4p}{(2\pi)^4} \log\det\left(
\begin{array}{cc}
\Pi_q & M_u \\ M_u^\dagger & \Pi_ u
\end{array}
\right) \ ,
\end{equation}
where, as shown in Eqs.~(\ref{eq-pi1}-\ref{eq-funPi}), the SU(3)$_c$ and SU(2)$_L$ indices of $\Pi_q$, $\Pi_u$ and $M_u$, allow us to express them as $6\times 6$, $3\times 3$ and $6\times 3$ matrices, respectively.

Since the SM fermions are in full representations of SU(3)$_L\times$SU(3)$_R\times$SU(2)$_L\times$U(1)$_Y$, the one-loop potential can spontaneously break this symmetry. Expanding to fourth-order in powers of the NGB fields $\Pi_5$ and $\Pi_6$, the most general potential is:
\begin{equation}\label{vf1}
V_f(\Pi_5,\Pi_6)=\frac{1}{2}m_{f, 5}^2\Pi_5^2+\frac{1}{2}m_{f, 6}^2\Pi_6^2+\lambda_{f, 5}\Pi_5^4+\lambda_{f, 6}(\Pi_6^4)+\lambda'_{f, 6}(\Pi_6^4)'+\lambda_{f, 56}\Pi_5^2\Pi_6^2 + \dots\ ,
\end{equation}
where the ellipsis stands for higher order terms. The indices of the NGB fields in the different products are contracted to form all the invariants allowed by the group structure, namely:
\begin{align}
& \Pi_5^2=\tr[\Phi_5^2]=\sum_a(\Pi_5^{\hat a})^2  \ , 
\qquad\qquad \Pi_5^4=\tr^2[\Phi_5^2]=[\sum_a(\Pi_5^{\hat a})^2]^2  \ , \\
& \Pi_6^2=\tr[\Phi_6\Phi_6^\dagger]=1/2\sum_b(\Pi_6^{\hat b})^2  \ , 
\qquad (\Pi_6^4)=\tr[\Phi_6\Phi_6^\dagger\Phi_6\Phi_6^\dagger]  \ , 
\qquad (\Pi_6^4)'=\tr^2[\Phi_6\Phi_6^\dagger]  \ .
\end{align}
The quadratic and quartic coefficients can be expressed as integrals of the correlators of the effective theory. They depend on the representation chosen for the fermions.

For negative $m^2_{f,5}$ and $m^2_{f,6}$ and suitable quartic couplings, this potential can trigger spontaneous symmetry breaking of H$_5$ and H$_6$.

Since the gauge interactions explicitly break G$_6$ to SU(3)$_V$, it is necessary to decompose the representation of $\Pi_6$ under this subgroup. Given that $(\bar 3,3)\sim 8+8+1+1$, we define: $\Phi_{(\bar 3,3)}\sim\phi_8+I_3\phi_1$, where $\phi_8$ and $\phi_1$ are complex fields, and $\phi_1$ can be identified with $h_6e^{i\theta_6}$. In terms of these fields, the potential generated by virtual exchange of elementary gluons at one-loop is:
\begin{equation}\label{vG}
V_G(\Pi_6)=\int \frac{d^4p}{(2\pi)^4} \log\det \Pi_G \ ,
\end{equation}
where $\Pi_G$ is an $8\times 8$ matrix, as defined in Eq.~(\ref{eq-pig1}).

Expanding to fourth order in $\Pi_6$:
\begin{align}
V_G(\Pi_6)=&m_{g,1}^2\phi_1\phi_1^\dagger+m_{g,8}^2\tr(\phi_8\phi_8^\dagger)+\lambda_{g,1}\tr(\phi_8\phi_8^\dagger\phi_8\phi_8^\dagger)+\lambda_{g,2}\tr^2(\phi_8\phi_8^\dagger)+\lambda_{g,3}\tr(\phi_8^\dagger\phi_8^\dagger)\tr(\phi_8\phi_8) \nonumber\\+&\lambda_{g,4}\tr(\phi_8\phi_8^\dagger)\phi_1\phi_1^\dagger+\lambda_{g,5}\tr(\phi_8^\dagger\phi_8^\dagger)(\phi_1)^2+\lambda_{g,6}\tr(\phi_8\phi_8)(\phi_1^\dagger)^2+\lambda_{g,7}(\phi_1\phi_1^\dagger)^2 + \dots\ ,
\end{align}
where the ellipsis stands for higher order terms. Matching the quadratic and quartic coefficients with the expansion of Eq.~(\ref{vG}) we obtain:
\begin{align}
&m_{g,1}^2=0 \ , \qquad \lambda_{g,7}=0 \ , \nonumber \\
&m_{g,8}^2=\frac{3}{f_6^2}\int_p\frac{2\Pi^G_{33}-\Pi^G_{81}-\Pi^G_{18}}{2Z_6+\Pi^G_{81}+\Pi^G_{18}} \ , \nonumber \\
&\lambda_{g,1}=-\frac{3}{2f_6^4}\int_p\frac{1}{(2Z_6+\Pi^G_{81}+\Pi^G_{18})^2}\left[(\Pi^G_{18})^2+3(\Pi^G_{33})^2+(\Pi^G_{81})^2+(\Pi^G_{33}-2\Pi^G_{11})(\Pi^G_{18}+\Pi^G_{81})\right.\nonumber\\
&\hskip6.5cm
\left. -\Pi^G_{18}\Pi^G_{81}+2Z_6(\Pi^G_{81}+\Pi^G_{18}+2\Pi^G_{11}-4\Pi^G_{33})\right] \ , \nonumber \\
&\lambda_{g,2}=-\frac{1}{2f_6^4}\int_p\frac{1}{(2Z_6+\Pi^G_{81}+\Pi^G_{18})^2}\left[(21\Pi^G_{33}-2\Pi^G_{11})(\Pi^G_{81}+\Pi^G_{18})-33(\Pi^G_{33})^2-5\Pi^G_{18}\Pi^G_{81}\right.\nonumber\\
&\hskip6.5cm
\left. +2Z_6(7\Pi^G_{81}+7\Pi^G_{18}-2\Pi^G_{11}-12\Pi^G_{33})\right] \ , \nonumber \\
&\lambda_{g,3}=-\frac{1}{4f_6^4}\int_p\frac{1}{(2Z_6+\Pi^G_{81}+\Pi^G_{18})^2}\left[6(\Pi^G_{18})^2+6(\Pi^G_{33})^2+6(\Pi^G_{81})^2+(2\Pi^G_{11}-33\Pi^G_{33})(\Pi^G_{18}+\Pi^G_{81})\right.\nonumber\\
&\hskip6.5cm
\left. +17\Pi^G_{18}\Pi^G_{81}+2Z_6(\Pi^G_{81}+\Pi^G_{18}-2\Pi^G_{11})\right] \ , \nonumber \\
&\lambda_{g,4}=-2\lambda_{g,5}=-2\lambda_{g,6}=\frac{6}{f_6^4}\int_p\frac{\Pi^G_{81}+\Pi^G_{18}-2\Pi^G_{33}}{2Z_6+\Pi^G_{81}+\Pi^G_{18}} \ . \label{eq-v-gauge}
\end{align}
Similar to the case of the fermions, we have shortened the notation of the representations by omitting the parenthesis. 

A few comments are in order. As expected, there is no contribution to the quadratic and quartic terms involving only the singlet. Besides, Eqs.~(\ref{vf1}) and~(\ref{eq-v-gauge}) are independent of $\theta_6$, which at this order remains as a NGB. In the specific realization of a two-site model (as well as simple extra dimensional models) $m_{g,8}^2\geq 0$, and the gauge potential does not misalign the vacuum. For $m_{g,8}^2>|m_{f,6}|^2$, only $\phi_1$ has a negative mass term and can spontaneously break H$_6$ to SU(3)$_V$.

\subsection{Symmetry breaking}\label{sec-sym-break}
We analyse in this section the potential for the singlets of SU(2)$_V$ and SU(3)$_V$: $h_5$ and $h_6$. For small $s_5$ and $s_6$ the potential can be expanded in powers of these variables; to fourth order it results in:
\begin{equation}\label{eq-vf-s-gral}
V(s_5,s_6)\simeq \alpha_5 s_5^2+\alpha_6 s_6^2+\beta_5 s_5^4+\beta_6 s_6^4+\beta_{56} s_5^2s_6^2 \ ,
\end{equation}
where the coefficients $\alpha$ and $\beta$ scale as $f^4$, and are given by integrals of the correlators. As we will show in sec.~\ref{sec-model-building}, in general $\beta_6$ is suppressed compared with the other quartic couplings. This suppression will have deep implications for model building, as well as for the phenomenology of the scalars.

Minimizing for a non-trivial vacuum we obtain:
\begin{equation}\label{eq-vevs-4}
\epsilon_5=\left(\frac{\alpha_6\beta_{56}-2\alpha_5\beta_6}{4\beta_5\beta_6-\beta_{56}^2}\right)^{1/2}
\ , \qquad 
\epsilon_6=\left(\frac{\alpha_5\beta_{56}-2\alpha_6\beta_5}{4\beta_5\beta_6-\beta_{56}^2}\right)^{1/2} \ .
\end{equation}
For $\beta_{56}\gg\beta_5,\beta_6$ the minimum is not stable and in general one of the symmetries is not spontaneously broken. As we discuss in the next subsection, this behavior has important consequences for model building.

\subsection{Spectrum of neutral scalar states}\label{sec-scalarspectrum}
From Eqs.~(\ref{eq-vf-s-gral}) and~(\ref{eq-vevs-4}), trading $\alpha_5$ and $\alpha_6$ for $\epsilon_5$ and $\epsilon_6$, one can obtain the mass matrix of the excitations of $h_5$ and $h_6$. In the basis $(h_5,h_6)$ the squared mass matrix is:
\begin{equation}
M^2=\left(\begin{array}{cc}8\frac{\beta_5}{f_5^2}(1 - \epsilon_5)^2 \epsilon_5^2&\frac{4}{\sqrt{6}}\frac{\beta_{56}}{f_5f_6}\epsilon_5^2  \sqrt{1 - \epsilon_5^2} \, \epsilon_6^2  \sqrt{1 - \epsilon_6^2} \,\\\frac{4}{\sqrt{6}}\frac{\beta_{56}}{f_5f_6}\epsilon_5^2  \sqrt{1 - \epsilon_5^2} \,\epsilon_6^2  \sqrt{1 - \epsilon_6^2} \,&\frac{4}{3}\frac{\beta_6}{f_6^2}(1 - \epsilon_6)^2 \epsilon_6^2\end{array}\right) \ .
\end{equation}
There are different interesting limits to study:
\begin{itemize}
\item $\epsilon_5^2\ll 1$: the masses and mixing angles of the physical states are given by 
\begin{equation}\label{eq-limit-xi5}
m_1^2\simeq 2\frac{\epsilon^2_5}{f_5^2}\left(4\beta_5-\frac{\beta_{56}^2}{\beta_6}\right) \ ,
\qquad
m_2^2\simeq \frac{4}{3}\frac{\epsilon^2_6(1-\epsilon^2_6)}{f_6^2}\beta_6\ , 
\qquad 
\theta\simeq \sqrt{\frac{3}{2(1-\epsilon^2_6)}}\frac{\epsilon_5 f_6}{\epsilon_6 f_5}\frac{\beta_{56}}{\beta_6} \ .
\end{equation}
In this case $m_1$ and the mixing are suppressed by powers of the small parameter $\epsilon_5$. The usual Higgs can be approximately identified with $h_5$, and there is a new scalar state which is expected to be rather light also, since its mass is suppressed by $\epsilon^2_6(1-\epsilon^2_6)$ and by the smaller quartic $\beta_6$.
\item $\beta_6\ll\beta_5,\beta_{56}$: this case is particularly relevant because it is realized in some interesting models, according to the discussion of sec.~\ref{sec-model-building}. The masses and mixing angles of the physical states are given by
\begin{align}\label{eq-limit-beta6}
& m_1^2\simeq 4\beta_5\frac{\epsilon^2_5(1-\epsilon^2_5)}{f_5^2}(1+\sqrt{1+c}) \ , \nonumber \\
& m_2^2\simeq 4\beta_5\frac{\epsilon^2_5(1-\epsilon^2_5)}{f_5^2}(1-\sqrt{1+c})+\frac{2}{3}\beta_6\frac{\epsilon^2_6(1-\epsilon^2_6)}{f_6^2}\frac{1+\sqrt{1+c}}{\sqrt{1+c}} \ , \nonumber\\ 
& \theta\simeq \frac{1}{\sqrt{2}}\frac{\sqrt{c}}{\sqrt{1+c}-\sqrt{1+c}} \ , \nonumber \\
& c=\frac{1}{6}\ \frac{\epsilon^2_6(1-\epsilon^2_6)\beta_{56}^2}{\epsilon^2_5(1-\epsilon^2_5)\beta_{5}^2}\ \frac{f_5^2}{f_6^2} .
\end{align}
It is worth analysing this result in some detail. Since the first term of $m_2^2$ is negative, the second term hast to be greater that the first one to obtain a stable minimum. However the second term is subdominant compared to the first one, in the expansion of small $\beta_6$. In this way one can see the tension introduced by the presence of a small $\beta_6$ and the requirement of spontaneous breaking of H$_5$ and H$_6$ by the potential. This scenario in general leads to $m_2$ smaller than $m_1$, due to the suppression of $\beta_6$ compared with $\beta_5$ and due to the tuning required to obtain a suitable minimum.

\item If in the previous case we consider $\beta_{56}\ll\beta_5$, we obtain
\begin{align}\label{eq-limit-beta6-beta56}
m_1^2\simeq 8\beta_5\frac{\epsilon^2_5(1-\epsilon^2_5)}{f_5^2} \ , 
\qquad m_2^2\simeq \frac{4}{3}\beta_6\frac{\epsilon^2_6(1-\epsilon^2_6)}{f_6^2} \ , 
\qquad \theta\simeq \frac{1}{2\sqrt{6}}\ \sqrt{\frac{1-\epsilon^2_6}{1-\epsilon^2_5}} \ \frac{\beta_{56}}{\beta_{5}} \ \frac{\epsilon_6 f_5}{\epsilon_5 f_6} \ .
\end{align}
This case is interesting because it is realized in several regions of the parameter space of the model studied in sec.~\ref{sec-2-site}. Since the mixing angle is small, the physical states can be approximated by $h_5$ and $h_6$. The mass of $h_5$ is controlled by the quartic $\beta_5$ and suppressed by $\epsilon^2_5$, whereas the mass of $h_6$ is suppressed by $\beta_6$, that was assumed to be smaller than the other quartic couplings, and by $\epsilon^2_6(1-\epsilon^2_6)$.
\end{itemize}

The previous examples show that in general one expects the presence of two light scalar states. In the case of small mixing angle, there is a state with quantum numbers similar to the SM Higgs boson, whose mass is suppressed by $\epsilon^2_5$, similar to the MCHM. Compared with that scenario, there is a new state, whose mass is in general suppressed by the small quartic $\beta_6$ and by $\epsilon^2_6(1-\epsilon^2_6)$.

\subsection{Model building}\label{sec-model-building}
In this subsection we discuss the impact of the fermion representations under G$_5$ on the potential $V(s_5,s_6)$. 
Using the form factors of Eq.~(\ref{eq-corr-f-phys}), and considering only the singlet scalar fields, the potential generated by exchange of virtual fermions at one-loop is:
\begin{equation}\label{eq-vf-s}
V_f(s_5,s_6)=-2N_c\int_p[\log(p^2\Pi_{u_L}\Pi_{u_R}-|\Pi_{u_Lu_R}|^2)+\log p^2\Pi_{d_L}] \ .
\end{equation}

By expanding $V_f$ of Eq.~(\ref{eq-vf-s}) in powers of $s_5$ and $s_6$ to fourth order it is possible to express the quadratic and quartic coefficients of Eq.~(\ref{eq-vf-s-gral}) in terms of integrals of the form factors. A simplified description of $V_f$ can be obtained by expanding also in powers of $1/Z_f$, or equivalently in powers of the fermion mixing $\epsilon_L$ and $\epsilon_R$. As discussed in Ref.~\cite{Panico:2012uw}, although for the third generation the mixing can be large, it can still provide a good estimate. Moreover, by performing numerical calculations to all orders in these parameters, we have checked that this expansion correctly describes the size of the coefficients.

Expanding in powers of $s_5$, $s_6$ and $1/Z_f$, we have analysed models with fermions in representations 1, 4, 5, 10, 14 and 16 of G$_5$. We find that: in all the cases $\alpha_5,\alpha_6$ and $\beta_{56}$ are of ${\cal O}(Z_f^{-1})$, whereas $\beta_{6}$ is of ${\cal O}(Z_f^{-2})$. For models involving only representations 1 and 5 $\beta_{5}$ is of ${\cal O}(Z_f^{-2})$, for example $\psi_q\sim 5$ and $\psi_u\sim 5$ or 1. For models where at least one of the fermions, $\psi_q$ or $\psi_u$, is in the representation 4, 10, 14 or 16, $\beta_{5}$ is of ${\cal O}(Z_f^{-1})$. If $P_{LR}$-symmetry is added to protect the $Zb_L\bar b_L$ coupling, representation 10 also gives $\beta_{5}$ of ${\cal O}(Z_f^{-2})$. As shown in sec.~\ref{sec-sym-break}, for $\beta_{56}\gg\beta_5,\beta_6$ it is not possible to break both symmetries simultaneously, thus models involving only representations 1, 5 and 10 with $P_{LR}$-symmetry are not favored. By considering the 2-site models of sec.~\ref{sec-2-site} we have computed the potential to all orders in $s_5$, $s_6$ and $Z_f$, and we have checked that this is indeed the situation. Therefore, although the mixings of the quarks of the third generation are large, the simplified analysis allows for a correct selection of the best representations. 

As is well known, the choice of the representation also affects flavor physics. In theories with a pNGB Higgs, fermion representations allowing more than one invariant Yukawa coupling induce Higgs mediated flavor violating processes that in general are incompatible with the current bounds from flavor physics for $f\sim$TeV~\cite{Agashe:2009di}. In order to proceed with our analysis, we will assume that, for a given type of fermion $ \psi_f$, all the generations are in the same representation. Once the representations for $\psi_q$ and $\psi_u$ are chosen, it is straightforward to count the number of invariants. As done in Eq.~(\ref{eq-leff-f}), a simple procedure is to decompose the representations of G$_5\times$G$_6$ in irreducible representations of H$_5\times$H$_6$: $\psi_f\sim\oplus_{rs}(\psi_f)_{rs}$, and count how many representations $rs$ in the decomposition of $\psi_q$ coincide with the ones in $\psi_u$. This number gives the number of independent invariants corresponding to Yukawa structures. There is one exception to this rule: if $\psi_q$ and $\psi_u$ are in the same representation of the group, there is one linear combination of the invariants that is independent of the pNGB, thus in this case one has to subtract one unit to the previous counting~\cite{Mrazek:2011iu}. A similar situation holds for the down sector and leptons. Taking into account these results, as well as the results from the previous paragraph, an interesting set of representations is: $\psi_q\sim 14$, $\psi_u\sim 1$ and $\psi_d\sim 10$ of SO(5). In this case there is just one non-trivial invariant for the Yukawa of the up-sector and one for the down-sector. Calling the invariants $I_{rs}$, where $r$ labels the representation of SO(4) and $s$ the representation of SU(3)$_L\times$SU(3)$_R$, the invariant for the up-sector can be symbolically written as: $I_{(1,1),(3,1)}-I_{(1,1),(1,3)}$, and the one for the down-sector as: $I_{(2,2),(3,1)}-I_{(2,2),(1,3)}$. In this counting we have taken into account that, since $q_L$ and $q_R$ are in the same representation of G$_6$, there is only one linear combination of invariants of H$_6$, that we have chosen as: $I_{r(3,1)}-I_{r(1,3)}$.

\noindent {\bf Model 14-1}

\noindent According to the discussion of the previous paragraph, we consider in detail the potential generated by fermions in the following representations of SO(5): $\psi_q\sim 14$ and $\psi_u\sim 1$, we will refer to this case as $14-1$. Under H$_5$ $\psi_q$ decomposes as: $14\sim(3,3)\oplus(2,2)\oplus(1,1)$, whereas $\psi_u$ is a singlet (1,1). These decompositions indicate the values that the subindex $r$ can take in the expressions for the form factors defined in Eqs.~(\ref{eq-pi1}) and~(\ref{eq-pi2}). For the representations of H$_6$: $s=(3,1),(1,3)$. As done previously, to shorten the notation we will omit the parenthesis, thus we will write $\Pi^f_{ijk\ell}$, with $i,j$ labeling the representation of H$_5$, and $k,\ell$ that of H$_6$, and similar for $M^f_{ijk\ell}$. We obtain: 
\begin{align}
&m_{f,5}^2=-\frac{2}{f_5^2}\int_p \frac{3}{4}\frac{5\Pi^q_{1131}-14\Pi^q_{2231}+9\Pi^q_{3331}}{Z_q+\Pi^q_{2231}}\ , \nonumber\\
&m_{f,6}^2=-\frac{2}{f_6^2}\int_p \left[\frac{\Pi^q_{2231}-\Pi^q_{2213}}{Z_q+\Pi^q_{2231}}+\frac{\Pi^u_{1131}-\Pi^u_{1113}}{Z_u+\Pi^q_{1113}}\right]
\ ,\nonumber\\
&\lambda_{f,5}=-\frac{2}{f_5^4}\int_p \frac{-3}{16(Z_q+\Pi^q_{2231})^2}[75 (\Pi^q_{1131})^2-140 \Pi^q_{1131} \Pi^q_{2231}+150 \Pi^q_{1131} \Pi^q_{3331}+44 (\Pi^q_{2231})^2\nonumber\\& \hskip3cm-252 \Pi^q_{2231} \Pi^q_{3331}+123 (\Pi^q_{3331})^2+16 Z_q (10 \Pi^q_{1131}-19 \Pi^q_{2231}+9
   \Pi^q_{3331})] \ ,\nonumber\\
&\lambda_{f,6}=-\frac{2}{f_6^4}\int_p \left[\frac{(\Pi^q_{2231}-\Pi^q_{2213})(2Z_q+3\Pi^q_{2213}-\Pi^q_{2231})}{12(Z_q+\Pi^q_{2231})^2}+\frac{\Pi^u_{1113}-\Pi^u_{1131}}{12(Z_u+\Pi^u_{1113})}-\frac{(\Pi^u_{1113}-\Pi^u_{1131})^2}{8(Z_u+\Pi^u_{1113})^2}\right] \ ,\nonumber\\
&\lambda'_{f,6}= 0 \ , \nonumber\\
&\lambda_{f,56}=-\frac{2}{f_5^2f_6^2}\int_p \frac{1}{32(Z_q+\Pi^q_{2231})^2}[5\Pi^q_{1113} \Pi^q_{2231}-5 \Pi^q_{1131} \Pi^q_{2213}-9 \Pi^q_{2213} \Pi^q_{3331}+9 \Pi^q_{2231} \Pi^q_{3313} \nonumber \\& \hskip4cm+Z_q (5 \Pi^q_{1113}-5 \Pi^q_{1131}-14 \Pi^q_{2213}+14 \Pi^q_{2231}+9 \Pi^q_{3313}-9 \Pi^q_{3331})] \ .
\end{align}

By keeping only the scalar fields $h_5$ and $h_6$, and expanding in powers of $s_5$ and $s_6$, one can also compute the potential $V(s_5,s_6)$ defined in Eq.~(\ref{eq-vf-s-gral}). For large $Z_f$, the leading non-trivial contribution to the coefficients of the potential are:
\begin{align}
&\alpha_5\simeq\frac{1}{2Z_q}\int_p(-9\Pi^q_{3331}+14\Pi^q_{2231}-5\Pi^q_{1131}) \ ,\nonumber\\ 
&\alpha_6\simeq\int_p\left[\frac{4}{Z_q}(\Pi^q_{2231}-\Pi^q_{2213})-\frac{2}{Z_u}(\Pi^u_{1131}-\Pi^u_{1113})\right] \ ,\nonumber\\ 
&\beta_5\simeq\frac{1}{2Z_q}\int_p(3\Pi^q_{3331}-8\Pi^q_{2231}+5\Pi^q_{1131}) \ ,\nonumber\\ 
&\beta_6\simeq\int_p\left[-\frac{1}{Z_q^2}(\Pi^q_{2231}-\Pi^q_{2213})^2+\frac{1}{2Z_u^2}(\Pi^u_{1131}-\Pi^u_{1113})^2\right] \ ,\nonumber\\ 
&\beta_{56}\simeq\frac{1}{2Z_q}\int_p(9\Pi^q_{3331}-9\Pi^q_{3313}-14\Pi^q_{2231}+14\Pi^q_{2213}+5\Pi^q_{1131}-5\Pi^q_{1113}) \ .
\end{align}
A few observations can be made: as already discussed, in this model all the coefficients are of ${\cal O}(Z_f^{-1})$, except for $\beta_6$ that is of ${\cal O}(Z_f^{-2})$; the sign of the quadratic coefficients is not fixed, thus for suitable regions of the parameter space it is possible to break both symmetries.

\section{A 2-site model}\label{sec-2-site}
Moose models can provide an effective description of a strongly coupled sector weakly coupled to external sources~\cite{Son:2003et}, they can also describe the lower level of resonances of extra-dimensional theories~\cite{Contino:2006nn}. In this section we consider the simplest case given by 2-site models; by using this framework one can explicitly compute the potential as well as some relevant observables. 

We follow the approach proposed in Ref.~\cite{DeCurtis:2011yx} to describe NGB fields arising from the SCFT (see also~\cite{Andres:2015oqa}). The setup is given by two sites containing different sets of fields, and connected by a set of non-linear sigma model fields. The first site, also called site-0, contains the same field content and symmetries as the SM but without scalar fields. For convenience, as already described in sec.~\ref{sec-extended-sym}, we add spurious fields to extend the gauge symmetry of the SM to G$_{5,0}\times$G$_{6,0}$, with $\rm{G}_5=\rm{SO}(5)\times U(1)_X$ and $\rm{G}_6=\rm{SU}(6)$, the subindex labelling the site-0. The corresponding gauge couplings are: $g_{X,0}$, $g_{5,0}$ and $g_{6,0}$. Following the discussion of sec.~\ref{sec-model-building}, we put $\psi_q$, $\psi_u$ and $\psi_d$ in the representations $14_{2/3}$, $1_{2/3}$ and $10_{2/3}$ of G$_5$, respectively. We take the fundamental representation of $G_6$ for the quarks, with $q_L$ in $(3,1)_{1/2\sqrt{3}}$ and $q_R$ in $(1,3)_{-1/2\sqrt{3}}$ of the subgroup H$_6$, defined in the next paragraph. 

In the second site, also called site-1, we put the fields that describe the first level of resonances of the SCFT. The spin one resonances can be described by the fields of the gauge symmetry G$_{5,1}\times$G$_{6,1}$, with gauge couplings $g_{X,1}$, $g_{5,1}$ and $g_{6,1}$. We add two sets of NGB scalar fields, $\Pi_5^1$ and $\Pi_6^1$,~\footnote{The index 1 indicates that these fields belong to site-1, and allows us to distinguish them from the non-linear sigma model fields connecting both sites that will be described below.} respectively describing the spontaneous breaking $\rm{G}_{5,1}/\rm{H}_{5,1}$ and $\rm{G}_{6,1}/\rm{H}_{6,1}$, by the strong dynamics, with $\rm{H}_5=\rm{SO}(4)\times\rm{U}(1)_X$ and $\rm{H}_6=\rm{SU}(3)_L\times\rm{SU}(3)_R\times\rm{U}(1)$. Each spontaneous breaking is characterized by a scale $f_{5,1}$ and $f_{6,1}$; both scales are assumed to be of the same order. Thus we include in the Lagrangian of site-1 the kinetic terms of the gauge fields and the NGB scalars, that as usual can be expressed in terms of the Cartan-Maurer form.

In addition, there are some fermion multiplets: $\Psi_q$, $\Psi_u$ and $\Psi_d$, that are in correspondence with the elementary ones, and are in the same representations of G$_5\times$G$_6$ as the elementary fermions. The fermions of site-1 are vector-like, with masses arising from the strong dynamics. The non-linear transformation properties of the NGB fields allow us to write gauge invariant Yukawa interactions with the fermions on site-1, as shown below:
\begin{align}\label{eq-l1}
{\cal L}_{1}\supset&\sum_{f=q,u,d}\bar\Psi_f (i\Dslash-m_{f,1})\Psi_f+
 f_1 \sum_{f=u,d}\sum_{rs} y^f_{rs}\overline{(U_5^\dagger U_6^\dagger \Psi_{qL})}_{rs}\ (U_5^\dagger U_6^\dagger \Psi_{fR})_{rs} + {\rm h.c.} \ ,
\end{align}
where $y^f_{rs}$ are dimensionless Yukawa couplings, and $f_1$ is a dimensionful parameter of the same order as the NGB decay constants. 
The second term of Eq.~(\ref{eq-l1}) generates Yukawa couplings for the elementary fermions after interactions between both sites. The representations chosen for the fermions allow for just one G$_5$ invariant for the Yukawa operator of the up sector, obtained by taking $r=(1,1)$. Since a $10_{2/3}$ of G$_5$ decomposes under H$_5$ as: $10_{2/3}\sim (2,2)_{2/3}\oplus(3,1)_{2/3}\oplus(1,3)_{2/3}$, there is also just one G$_5$ invariant for the down sector, obtained by taking $r=(2,2)$. The index $s$ takes the values (3,1) and (1,3). Since for $y^f_{r31}=y^f_{r13}$ the Yukawa is independent of $\Pi_6$, the linear combination $y^f_{r31}-y^f_{r13}$ measures the Yukawa coupling involving $\Pi_6$, whereas $y^f_{r31}+y^f_{r13}$ leads to a mixing between $\Psi_{q}$ and $\Psi_{f}$. Thus in the present 2-site model the Yukawa couplings are proportional to $y^f_{r31}-y^f_{r13}$. These results are in agreement with the arguments of sec.~\ref{sec-model-building} for the low energy effective theory.

In Eq.~(\ref{eq-l1}) we have included only one chiral structure, avoiding terms with the opposite chiral structures: $\overline{(U^\dagger \Psi_{qR})}_{rs}\ (U^\dagger \Psi_{fL})_{rs}$. As discussed in Ref.~\cite{Carena:2014ria}, there is no fundamental reason for this, the motivation being that in this way one can obtain a finite one-loop potential. 

All the couplings of site-1, generically called $g_1$, are assumed to be larger than the SM ones but still perturbative: $g_{SM}\ll g_1\ll 4\pi$. All the dimensionful parameters are assumed to be of order TeV, the masses of the fermions being $m_{f,1}\sim g_1 f_1$.

The two sites are connected by a set of non-linear sigma models: $\Pi_{X,0}$, $\Pi_{5,0}$ and $\Pi_{6,0}$, associated to the gauge symmetry groups. These fields parametrize the cosets $(\rm{G}_{5,0}\times \rm{G}_{5,1})/\rm{G}_{5,0+1}$ and $(\rm{G}_{6,0}\times \rm{G}_{6,1})/\rm{G}_{6,0+1}$, where $\rm{G}_{5,0+1}$ and $\rm{G}_{6,0+1}$ are the diagonal subgroups. 
The field $U_{5,0}=e^{i\sqrt{2}\Pi_{5,0}/f_{5,0}}$ transforms linearly under $\rm{G}_{5,0}\times \rm{G}_{5,1}$, as: $U_{5,0}\to \hat g_{5,0}U_{5,0}\hat g_{5,1}^\dagger$, with $\hat g_{5,0}\in \rm{G}_{5,0}$ and $\hat g_{5,1}\in \rm{G}_{5,1}$. Similar properties apply to $U_{6,0}=e^{i\sqrt{2}\Pi_{6,0}/f_{6,0}}$. Besides, these fields play an important role in the realization of partial compositeness, that is obtained by linear interactions between the fermions on site-0 and site-1, through the following terms:
\begin{equation}
{\cal L}_{\rm{mix}}\supset f_0 \sum_{f=q,u,d} \lambda_f \bar \psi_f U_{X,0}^{2/3}U_{5,0}U_{6,0}\Psi_f + \rm{h.c.}
\end{equation}
with $f_0$ a dimensionful constant and $\lambda_f$ a dimensionless coupling that controls the size of the mixing between the fermions. All the dimensionful parameters, namely: the decay constants and $f_0$, are taken of order TeV, whereas the couplings $\lambda_f$ can be hierarchical. 

Let us briefly count the number of NGB fields. One can choose a gauge where the fields $\Pi_{X,0}$, $\Pi_{5,0}$ and $\Pi_{6,0}$ are removed. In this gauge it is straightforward to see that the gauge fields of the corresponding cosets become massive, with masses $m_1\sim g_1 f_0/\sqrt{2}$, whereas the gauge fields of the diagonal cosets remain massless. The NGBs $\Pi_{5,1}$ and $\Pi_{6,1}$ remain in the spectrum; $\Pi_{5,1}$ can be associated with the Higgs multiplet and $\Pi_{6,1}$ gives a new scalar multiplet. The previous gauge is not the unitary one, that can be obtained by demanding the absence of mixing terms between the gauge and the NGB fields, as described in Refs.~\cite{Panico:2015jxa} and~\cite{Alvarez:2016ljl}. The decay constants of the physical NGB fields $\Pi_5$ and $\Pi_6$ are given by: $f^{-2}_5=f^{-2}_{5,0}+f^{-2}_{5,1}$ and $f^{-2}_6=f^{-2}_{6,0}+f^{-2}_{6,1}$, respectively.

The degree of compositeness of the light fermions is given by $\epsilon_f=\lambda_f/g_1$, where we have considered $f_0\sim f_1$. The light fermions are mostly elementary if $\epsilon_f\ll 1$, leading to small mixing, whereas for $\epsilon_f\sim 1$ the mixing between the fermions on site-0 and site-1 is large, leading to partially composite states. We are interested in the anarchic scenario, that in the 2-site model is obtained by taking structureless Yukawa couplings, with all the elements of the matrices being of the same order: $(y^f_{rs})_{jk}\sim g_1$, where $j$ and $k$ are flavor indices. Keeping the NGB dependence, the mass matrices of the SM fermions are:
\begin{equation}\label{eq-mjk}
(m_f)_{jk}\sim \frac{\sqrt{5}}{2} \epsilon_{qj}\epsilon_{fk} s_5 c_5 s_6 c_6 f_1(y^f_{rs})_{jk} \ ,
\end{equation}
thus a light fermion requires at least the mixing of one of the chiralities to be small, whereas the top requires both mixings of order one to be able to obtain a Yukawa coupling $y_t\sim 1$. Besides the masses, it is also necessary to reproduce the mixing angles of the CKM matrix. Taking hierarchical mixings $\epsilon_f$ leads to the hierarchies between the masses and CKM angles.~\footnote{Although hierarchical $\lambda_{fj}$ may look arbitrary, they appear naturally when considering the running of the couplings between the elementary fermions and the SCFT~\cite{Contino:2004vy,Panico:2015jxa}. A simple realization can be obtained in extra dimensional models~\cite{Grossman:1999ra,Gherghetta:2000qt}.} 
Following Refs.~\cite{Csaki:2008zd,Agashe:2008uz} we take:
\begin{equation}\label{anarchy1}
\frac{\epsilon_{q1}}{\epsilon_{q2}}\sim \lambda_C\ , \qquad
\frac{\epsilon_{q2}}{\epsilon_{q3}}\sim \lambda_C^2 \ , \qquad
\frac{\epsilon_{q1}}{\epsilon_{q3}}\sim \lambda_C^3 \ .
\end{equation}
where $\lambda_C$ is the Cabibbo parameter. By using $v_{SM}=\epsilon_5 f_5$, $m_t\sim v_{SM}/\sqrt{2}$ and $y^u_{rs}\sim g_1$, the following relations arise from anarchy:
\begin{align}
&\epsilon_6 \sqrt{1 - \epsilon_6^2} \, \epsilon_{q3}\epsilon_{u3}g_1\sim1 \ , & \epsilon_{d3}\sim \epsilon_{u3}\frac{y^u}{y^d}\frac{m_b}{m_t} \ , \nonumber \\
&\epsilon_{u2}\sim\frac{\epsilon_{u3}}{\lambda_C^2}\frac{m_c}{m_t} \ , & \epsilon_{d2}\sim \frac{\epsilon_{u3}}{\lambda_C^2}\frac{y^u}{y^d}\frac{m_s}{m_t} \ , \nonumber \\ 
&\epsilon_{u1}\sim\frac{\epsilon_{u3}}{\lambda_C^3}\frac{m_u}{m_t} \ , & \epsilon_{d1}\sim \frac{\epsilon_{u3}}{\lambda_C^3}\frac{y^u}{y^d}\frac{m_d}{m_t} \ .
\label{anarchy2}
\end{align}
The factor $y^u/y^d$ in the second column is of ${\cal O}(1)$ if all couplings on site-1 are of the same order, but differs from this estimate if the Yukawa couplings of the down- and up-sectors on site-1 are taken different. The upper-left equation of~(\ref{anarchy2}) shows that $\epsilon_6 \sqrt{1 - \epsilon_6^2}$ cannot be too small if $g_1$ is required to be perturbative since in the most favorable case, for $\epsilon_{q3}\sim\epsilon_{u3}\sim 1$, it leads to: $g_1\sim(\epsilon_6 \sqrt{1 - \epsilon_6^2})^{-1}$.

Integrating out the spin-one fields and fermions on site-1, one obtains the effective theory described in sec.~\ref{sec-extended-sym}. The correlators of Eqs.~(\ref{eq-leff-f}) and~(\ref{eq-leff-g}) can be computed explicitly in terms of the parameters of the 2-site theory. We show them in Ap.~\ref{app-correlators}.

\subsection{Dominant contributions to $\epsilon_K$}\label{sec-epsK}
Integrating out the heavy resonances, new contributions to the Wilson coefficients $C_i$ of dimension-6 operators contributing to $K-\bar K$ mixing are generated. The most dangerous contributions are those generated by exchange of massive gluons, that give a contribution to $C_4$:
\begin{equation}
C_4=C_4^{SU(3)}\epsilon^2_6 (1 - \epsilon^2_6)\frac{f_{6,1}^2}{2(f_{6,0}^2+f_{6,1}^2)} \ ,
\end{equation}
where $C_4^{SU(3)}$ is the contribution in the usual case with SU(3) global symmetry only.

Bounds on $C_4^{sd}$ from $\epsilon_K$ usually require $m_{cp}\gtrsim 10\div 30$ TeV. In the present case, taking $f_{6,0}\simeq f_{6,1}$, for $\epsilon_6 \sqrt{1-\epsilon_6^2}\sim 0.2\div 0.5$ the scale of the masses of the composite gluons can be lowered to $\sim 2.5$ TeV. Since $0<\epsilon_6<1$, $\epsilon_6 \sqrt{1-\epsilon_6^2}$ is always smaller than (at most equal to) 0.5.

As mentioned in sec. \ref{sec-eff-th}, the decomposition of the adjoint of G$_6$ under H$_6$ contains a $(1,1)_0$ - the generator of the U(1) factor in H$_6$. The presence of this state induces an extra contribution to $C_5$, that is not suppressed by $\epsilon_6$: 
\begin{equation}
C_5=C_4^{SU(3)}\frac{1}{6} \ .
\end{equation}
The contribution of $C_5^{sd}$ to $\epsilon_K$ is suppressed by $N_c$, and in the present case by an extra factor 1/6 compared with the usual case. Therefore the contribution of this state is not critical for~$\epsilon_K$.

\subsection{Estimation of the top mass}\label{sec-mtop}
The mass of the top can be approximated by the coefficient $j=k=3$ of Eq.~(\ref{eq-mjk}).
Taking $\epsilon_{q3}\sim\epsilon_{u3}\sim 1$, $f_1\simeq 1$ TeV and the values for $\epsilon_6$ that saturate $C_4^{sd}$, we obtain
\begin{equation}
y^u_{rs}\sim 2.4\div 0.6 \ ,
\end{equation}
where the largest value corresponds to $\epsilon_5\sim 0.3$ and $\epsilon_6\sim 0.2$ and the smallest one to $\epsilon_5\sim 0.5$ and $\epsilon_6 \sqrt{1 - \epsilon^2_6}\sim 0.5$.
Although there is an extra suppression from the factor $\epsilon_6 \sqrt{1 - \epsilon^2_6}$ compared with the usual case, ${\cal O}(1)$ Yukawa can lead to the top mass in the present model.

\section{Numerical results}\label{sec-num-results}
In this section we present some numerical results that arise from the 2-site model described in sec.~\ref{sec-2-site}. We focus first on the issue of spontaneous symmetry breaking of the extended symmetries: H$_5$ that contains the EW symmetry group, and H$_6$ that contains the color symmetry group. By performing a random scan over the parameters of the 2-site models, we have checked the properties of the potential discussed in sec.~\ref{sec-potential}. In particular we have checked the predictions of sec.~\ref{sec-model-building} for different combinations of the representations 1, 5, 10 and 14 of the fermions under SO(5), and their impact on the possibility to obtain successful breaking of both symmetries. We have found that, in models with fermions in combinations of the representations 1, 5, and 10, in general there is no spontaneous breaking at all, and in some cases only H$_5$ is broken.~\footnote{We imposed Left-Right parity for the 10 of SO(5).} We have also checked that when the 14 is included there are suitable regions of the parameter space where the symmetry is spontaneously broken to K$_5\times$K$_6$. 
\begin{figure}[!htpb]
  \centering
  \begin{tabular}{@{}p{.5\textwidth}@{\quad}p{.5\textwidth}@{}}
    \subfigimg[width=\linewidth]{\bf{1.A)}}{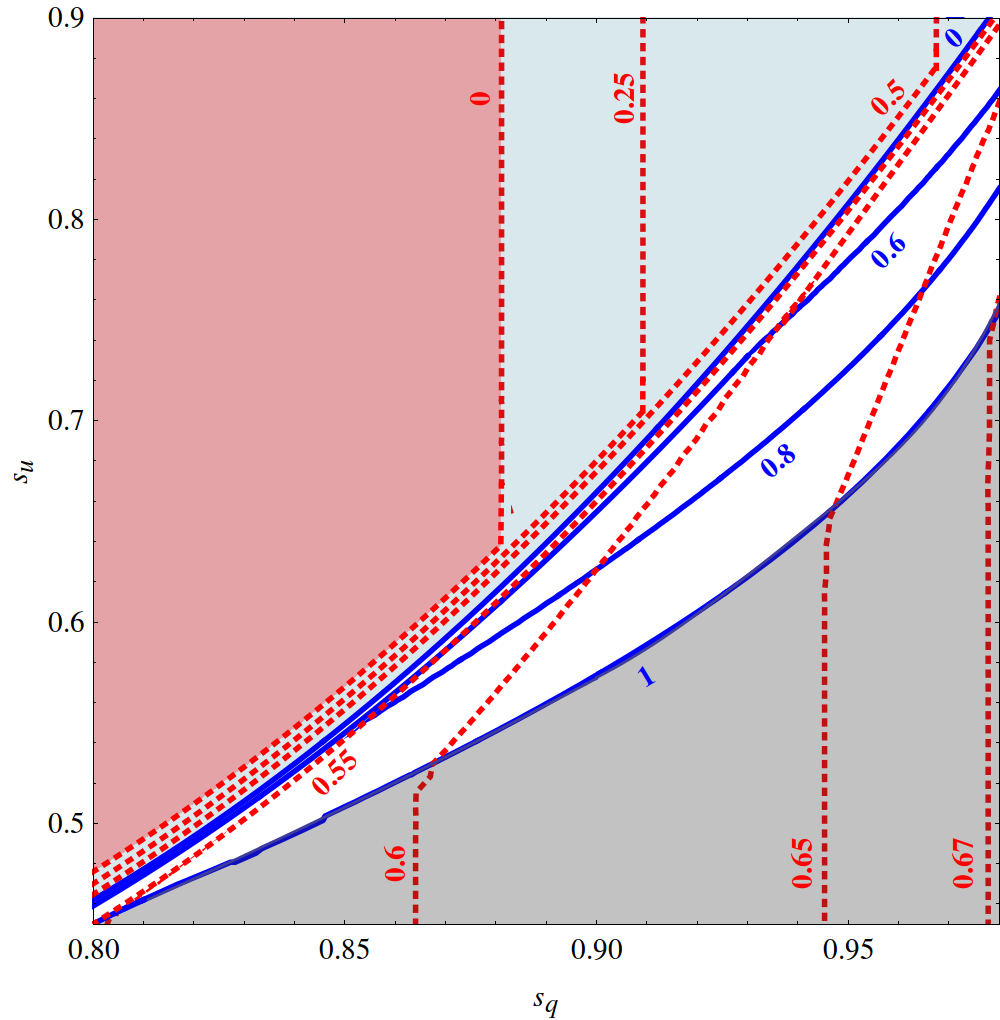} & \subfigimg[width=\linewidth]{\bf{1.B)}}{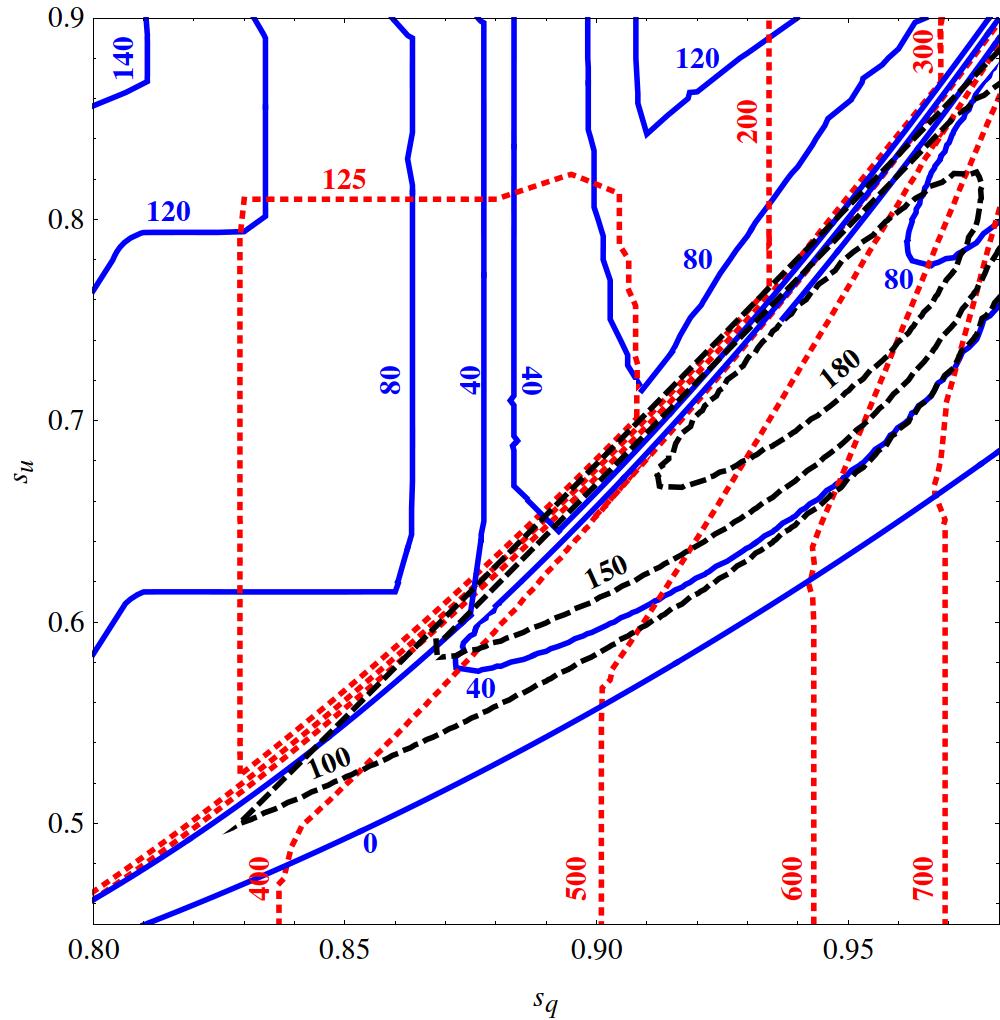} \\
    \subfigimg[width=\linewidth]{\bf{2.A)}}{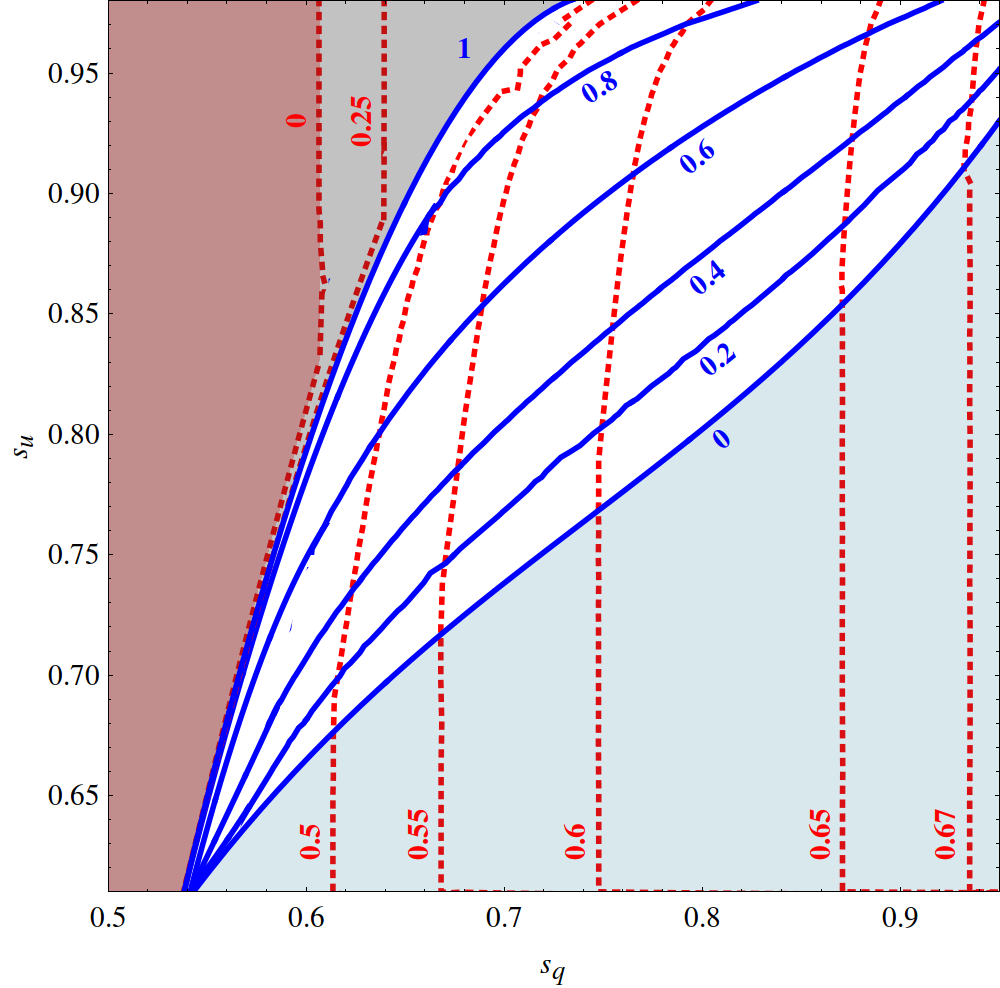} &  \subfigimg[width=\linewidth]{\bf{2.B)}}{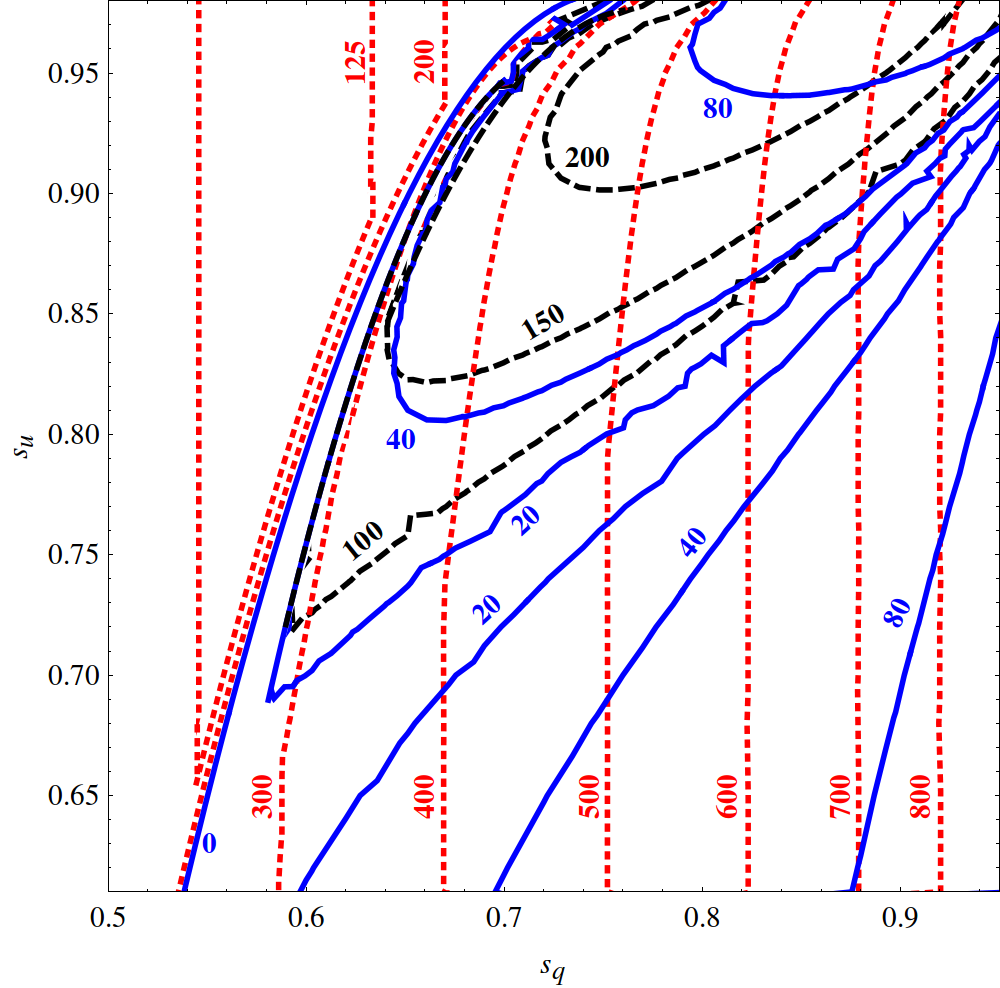}
  \end{tabular}
  \caption{Contour plots for the values of $\epsilon_5$ and $\epsilon_6$ at the minimum of the potential (red dotted and blue solid lines respectively in plots labeled with an A) and for the masses of the Higgs, the new scalar $h_6$ and the top quark (red dotted, blue solid and black dashed lines respectively in plots labeled B) as functions of the mixings $s_q$ and $s_u$ for two different sets of parameters, labeled with 1 and 2, as described in the text.. In the A plots, the gray region corresponds to $\epsilon_6 = 1$, the light blue region to $\epsilon_6 = 0$ and the red region to $\epsilon_5 = 0$. All masses are expressed in GeV.}\renewcommand\thefigure{\arabic{figure}}    
  \setcounter{figure}{0} \label{num_res_fig} 
\end{figure}

In the rest of this section we will present the results for model $14$-$1$ that, as discussed in sec.~\ref{sec-model-building}, is one of the most interesting models. We start by explaining our scan of the parameter space. It is useful to define mixing angles for the fermions by the following relation: $\tan\theta_f=\lambda_f f_1/m_f$, with $m_f$ the mass of the vector-like fermions on site-1. We will use $s_f$ and $c_f$ for $\sin\theta_f$ and $\cos\theta_f$ respectively; notice that with this parametrization the absence of mixing is given by $s_f=0$, whereas large mixing corresponds to $s_f\simeq 1$. We have fixed the masses $m_f$ and the Yukawa couplings $y^u_{rs}$, and we have varied $s_q$ and $s_u$. For the bosonic sector we have fixed the ratio between the couplings $g_0/g_1$ and we have matched the coupling of the diagonal subgroups with the SM gauge couplings. We have also fixed the masses of the spin-one resonances, $m_1$. In Fig. \ref{num_res_fig}  we show the results obtained when scanning over $s_q$ and $s_u$ for a point with parameters $g_0/g_1 = 0.26$, $m_q = m_u = 2$ TeV, $m_1 = 2.5$ TeV, $y^u_{1131} = 0.4$, $y^u_{1113} = -0.82$ and $f_5 = f_6 = 1$ TeV (upper panels, labeled $1$) and another with $g_0/g_1 = 0.25$, $m_q = 3$ TeV, $m_u = 1.4$ TeV, $m_1 = 3$ TeV, $y^u_{1131} = -0.7$, $y^u_{1113} = 0.5$ and $f_5 = f_6 = 1$ TeV (lower panels, labeled $2$). The panels on the left (labeled A) show the expectation values obtained for $s_5$ (red dotted) and $s_6$(blue solid), with the regions where $\epsilon_6 = 0$ colored in light blue, those were $\epsilon_5 = 0$ in red and the ones with maximal H$_6$ breaking ($\epsilon_6 = 1$) in gray. The white, uncolored regions in each A panel are the regions where we found spontaneous symmetry breaking to K$_5\times$K$_6$, and thus the most interesting phenomenologically. The panels labeled with a B show the masses of the Higgs boson (red dotted) and the $h_6$ scalar (blue solid). We have verified that the mixing angle between these scalars is very small ($\leq 0.1$ rad) and so their identification with the mass eigenstates is a reasonable approximation. It is worth noting that the mass of $h_6$ was always found to be smaller than that of the Higgs within the phenomenologically interesting region; this is a general feature of this model and a consequence of the discussion presented in sec. \ref{sec-scalarspectrum}. The phenomenological implications of having this additional light scalar will be analysed in the following section. In the B panels we also show the mass of the SM top quark (black dashed), which is calculated as the mass of the lightest up-type quark. Even though the vev's and Higgs masses shown in these plots are not completely realistic, more realistic values could be found with a more extensive scan. All sharp angles and jagged lines in these plots are only a consequence of the limitations of the numerical calculations employed and do not carry any physical meaning.

\section{Discussions}\label{sec-discussions}
\subsection{Neutron dipole moments}
Besides the constraints from $K-\bar K$ mixing, the electromagnetic (EDM) and chromomagnetic dipole moments (CDM) of the neutron give the most important set of constraints in models with anarchic flavor~\cite{Agashe:2004cp,Redi:2011zi,Delaunay:2012cz,Konig:2014iqa}. In theories where the SCFT has a global symmetry SU(3), gauged by the color interactions, assuming that the dipole operators are induced at loop level the following bound is obtained: $f\gtrsim 4.5$~TeV~\cite{Panico:2015jxa}. As expected, in the presence of the larger SU(6) symmetry, given the representations chosen for the fermions, the contribution to the Wilson coefficients of operators with $LR$ chiral structures require insertions of the scalar $\Pi_6$. In particular, the group structure associated to the SU(6) symmetry of the EDM operators is identical to the structure of the mass operators, therefore at least one insertion of the scalar field $\Pi_6$ is required for the EDM operator. Resumming these insertions and evaluating in the vacuum, the bound is relaxed as: $f/(\epsilon_6\sqrt{1-\epsilon_6^2})\gtrsim 4.5$ ~TeV. As an example, $f=2$~TeV is compatible with the bounds for $\epsilon_6\lesssim 0.2$ or $\epsilon_6\gtrsim 0.98$.

\subsection{Phenomenology of the scalar states}\label{sec-light-h}
An interesting and distinctive phenomenology of the present model is given by the presence of the new scalar states. Let us briefly characterize this sector of the theory. As discussed in sec.~\ref{sec-scalarspectrum}, there are two light scalar states, corresponding to the excitations of the fields that acquire a vev at loop level, $h_5$ and $h_6$. As mentioned in the previous section, we find that the mixing between these fields is small, and therefore, to leading order the mass eigenstates can be identified with $h_5$ and $h_6$. Since $h_5$ arises from SO(5)/SO(4), its interactions with the EW gauge bosons are similar to those of the Higgs in the MCHM. On the other hand, $h_6$ does not interact with the gauge bosons at tree level. The Yukawa couplings at low energies are shown in Eqs.~(\ref{eq-yf5-aprox}) and~(\ref{eq-yf6-aprox}). In the full theory the scalars also interact with the resonances.

Let us characterize the creation and decay of these scalars at the LHC. As for the usual Higgs, the main creation channel of the scalars at the LHC is gluon fusion, with much smaller contributions from creation in association with $t\bar t$, as well as creation in association with EW gauge bosons for the case of $h_5$. Gluon fusion is induced at loop level by virtual exchange of fermions $f^{(n)}$. In the limit of heavy fermions, $2m_f^{(n)}\gg m_{j}$, this coupling is proportional to $\sum_{n\geq0} y_{f,j}^{(n)}/m_f^{(n)}$, where $j=5,6$ labels the scalars and quantities with $n = 0$ pertain to the would be zero mode corresponding to the SM fermion. By standard algebraic manipulations it is possible to rewrite this expression as ${\rm tr}(Y_{f,j}\mathcal{M}_f^{-1})$, with $Y_{f,j}=\partial_{h_j}\mathcal{M}_f$ and $\mathcal{M}_f$ the fermion mass matrix of the full theory. As usual in models with NGB scalars, this trace leads to a compact expression:
\begin{align}
{\rm tr}(Y_{u,5}\mathcal{M}_u^{-1})=\frac{1}{v_{SM}}\frac{1-2\epsilon_5^2}{\sqrt{1-\epsilon_5^2}}\ \ , \label{eq-ggh5-aprox}
\\
{\rm tr}(Y_{u,6}\mathcal{M}_u^{-1})=\frac{1-2\epsilon_6^2}{\sqrt{6}f_6\epsilon_6\sqrt{1-\epsilon_6^2}}\ \ . \label{eq-ggh6-aprox}
\end{align}
The case of $h_5$ has been extensively studied, as well as its dependence on the fermion representations~\cite{Azatov:2011qy,Montull:2013mla,Carena:2014ria}. The coupling to $h_6$ has a similar behavior, but in terms of the parameters associated to G$_6$. By comparing with Eq.~(\ref{eq-yf6-aprox}), one can see that the sum over states is saturated by the lightest fermion, in this case corresponding to the top.~\footnote{The light quarks and the leptons do not contribute in the present setup, see Ref.~\cite{Carena:2014ria} for discussions.} Given these results, creation of $h_j$ by gluon fusion and $t\bar t h_j$ processes is similar to Higgs creation in the MCHM.

Since the couplings of $h_5$ are similar to the couplings of the Higgs in the MCHM, its decay width and branching ratios are similar to that case. On the other hand $h_6$ can decay at tree level to SM fermion-antifermion, with Yukawa coupling given by Eq.~(\ref{eq-yf6-aprox}). This Yukawa coupling can be written in terms of the usual SM Yukawa with the Higgs, up to a factor depending on $\epsilon_6$ and $f_6$: 
\begin{equation}\label{eq-ffh6-aprox}
y_{f,6}\simeq y_f^{SM}\ \delta\ , \qquad \delta=\frac{v_{SM}}{f_6}\frac{1-2\epsilon_6^2}{\sqrt{6}\epsilon_6\sqrt{1-\epsilon_6^2}} \ .
\end{equation}
At loop level, for a light $h_6$, the most important decay channels are pairs of gluons and photons. The amplitude of the former is modulated by Eq.~(\ref{eq-ggh6-aprox}). The amplitude of the later is modulated by a similar factor, multiplied by the corresponding factors of electric charge. Compared with $h_5$, in the one-loop level contribution to $h_6\gamma\gamma$ there is no contribution from virtual EW gauge bosons. For light $h_6$ we expect diphoton and $b\bar b$ final states to be the most relevant ones, whereas for a heavier scalar the final state $t\bar t$ will become important, and massive diboson states mediated by loops of fermions also will become available. 

According to the results of Eqs.~(\ref{eq-ggh6-aprox}) and (\ref{eq-ffh6-aprox}), the creation cross-section by gluon fusion and the partial widths of $h_6$ are similar to the ones of the SM Higgs, with a rescaling factor $\delta^2$, and changing the Higgs mass by the mass of $h_6$. The only qualitative difference being that decays to pairs of massive gauge bosons are not present at tree level for $h_6$. 

In Refs.~\cite{Aad:2014ioa} and~\cite{CMS-PAS-HIG-14-037}, the ATLAS and CMS Collaborations have presented limits in the production cross-section of new scalar states decaying to photon pairs. In order to compare with those bounds, we consider a reference mass for $h_6$ of 80 GeV. In this case the production cross-section of $h_6$ is: $\sigma\simeq 45 {\rm pb}\times \delta^2$, and the BR to photon pair is of order $10^{-3}$, leading to a bound $\delta\lesssim 1$. For $f_6= 1$ TeV, this bound can be satisfied as long as $0.1\lesssim\epsilon_6\lesssim 0.99$. Therefore a new scalar state, somewhat lighter than the Higgs, could have evaded detection at the LHC.

Besides the neutral scalar fields already discussed, the present model contains an axion-like state, associated to the spontaneous breaking of the U(1) subgroup of H$_6$. Indeed the vacuum expectation value $v_6$ spontaneously breaks this global U(1), and the scalar phase $\theta_6$ becomes the associated NGB. 
In order to understand the phenomenology of this state, let us briefly discuss its interactions. 
Due to the embedding of the quarks into the fundamental representation of G$_6$, the Left- and Right-handed quarks of the SM have opposite charges under U(1). The axion-like scalar has interactions with the fermions of the form: $\theta_6 h_5 \bar q_L\gamma^5 q_R$, that at one-loop level induce interactions $\theta_6 F_{\mu\nu}\tilde F^{\mu\nu}$ and $\theta_6 G_{\mu\nu}\tilde G^{\mu\nu}$ with photons and gluons. It is usually assumed that the axions acquire mass only through non-perturbative effects, however it has been shown recently that they can also obtain a mass at perturbative level~\cite{Ibanez:2015uok,Herraez:2016dxn}. The simplest scenario is obtained by considering the presence of a 3-form field, that does not propagate since its equation of motion fixes its field strength to be a constant. It has been shown that the interactions of this form with the axion induce a potential, with a mass for the axion. If the dimensional coupling is of order TeV, the axion mass can be of order several hundred GeV to a few TeV. This scalar state could mix with the other scalars. It could also be produced by gluon and photon fusion, and decay to pairs of photons, gluons and fermion-antifermions. A precise description of its phenomenology requires assumptions on the specific realization, since there can be several 3-forms involved. We will not consider those details in the present work.

The new scalar sector also contains octets of SU(3)$_c$. The mass of these multiplets can be estimated from $m_{g,8}^2$ in Eq.~(\ref{eq-v-gauge}), leading to $m^2\simeq\alpha_s/(4\pi) g_1^2 f_6^2 \sim {\cal O}({\rm TeV}^2)$. Being color octets, these states can be copiously produced by QCD interactions, with a large branching fraction to dijets. ATLAS and CMS have searched for narrow resonances in dijet final states~\cite{Aad:2014aqa,Khachatryan:2015sja,Sirunyan:2016iap}, giving bounds on the mass of the scalar octets of order 3 TeV.

\subsection{Tuning}
The tuning can be roughly estimated as $(\epsilon_5\epsilon_6)^{-2}$. The factor $\epsilon_5^{-2}$ is the usual tuning of the MCHM, that could be increased depending on the representations of the fermions~\cite{Panico:2012uw}. EW precision tests usually require $\epsilon_5^2\lesssim 0.1\div 0.3$, leading to a tuning of order $4\div 10$. Numerical calculations using different representations for the fermions show that the tuning varies between 5 and $10^3$ in the MCHM~\cite{Carena:2014ria}. The need in the present paper for the spontaneous breaking of another symmetry is expected to worsen the tuning, introducing an extra factor $\epsilon_6^{-2}$. As discussed in sec.~\ref{sec-epsK}, by demanding the masses of the resonances to be of order 2.5 TeV, constraints from the Kaon system give the bounds: $\epsilon_6\lesssim 0.2\div 0.98$. In the most stringent case of $\epsilon_6\lesssim 0.2$, we expect to increase the tuning by a factor of order 25, whereas in the most favorable case we expect no sizable rise of the tuning compared with the MCHM. Going beyond this general estimate requires considering the details of the model, in particular the embedding of the fermions, as well as numerical calculations.

Following the definition for the sensitivity parameter of Refs.~\cite{Barbieri:1987fn, Anderson:1994dz, Panico:2012uw}, we have computed the tuning in our model with the fermions $q_L$ and $u_R$ in the representations 14 and 1 of SO(5). For the regions of the parameter space of Fig.~\ref{num_res_fig} we find a tuning of order 50-300, with a few points where it rises to 400. One can compare this tuning with the results of Ref.~\cite{Carena:2014ria}, that has reported a tuning of order $80\div 300$ for the MCHM$_{14-1}$. However in that paper the Higgs, top and $W$ masses were restricted to their physical values, selecting a region of the parameter space with non-natural cancellations in the scalar potential and thus increasing the tuning. In Fig.~\ref{num_res_fig}, the masses of the Higgs, top and EW gauge bosons are not fixed to their physical values. Constraining these masses requires non-trivial cancellations also in the model with the extended symmetry G$_6$, and we expect a larger tuning in that case.

\subsection{Other representations and symmetry groups}
The need for a larger quartic coefficient $\beta_6$ suggests changing the representation of the quarks under G$_6$. A larger $\beta_6$ could improve the tuning of the model by allowing a larger region in the parameter space with the right spontaneous breaking of symmetries. Cancellation of $C_4$ requires $q_L\sim(3,1)$ and $q_R\sim(1,3)$ of SU(3)$_L\times$SU(3)$_R$, thus the quark representations under SU(6) must contain these representations of the subgroup SU(3)$_L\times$SU(3)$_R$. However we have not been able to find other small representations of SU(6) that satisfy those conditions.

We have also explored the possibility of embedding the SO(4) as well as the extended SU(3)$_L\times$SU(3)$_R$ symmetry groups into a single unified group. We have found several examples of groups that, after spontaneous breaking to SO(4)$\times$SU(3)$_L\times$SU(3)$_R$, deliver a NGB field transforming as a (2,2,3,3) of the unbroken subgroup, playing the role of the Higgs field in the extended quark sector~\cite{DaRold:2012sz}, the most interesting one being the exceptional group E$_8$. However we have not been able to obtain suitable representations for the quarks. For example, for $q_L$, adding custodial symmetry to protect the $Zb_L\bar b_L$ coupling, one would require a representation containing a (2,2,3,1) of the unbroken subgroup, whereas for $u_R$ one would require (1,1,1,3) or (1,3,1,3). By exploring the lowest dimensional representations of the unified groups, we have not been able to find the proper representations for the quarks.

\section{Conclusions}
Flavor anarchic composite Higgs models with partial compositeness of the SM fermions offer a rationale to understand the hierarchies in the flavor of the quarks and leptons. The most stringent constraint from flavor physics, arising from $\epsilon_K$ in $K-\bar K$ system, pushes the scale of compositeness to $f\sim{\cal O}(10)$TeV, introducing a little hierarchy problem for a light Higgs. Ref.~\cite{Bauer:2011ah} showed that if the composite sector has an SU(3)$_L\times$SU(3)$_R$ global symmetry, with SU(3)$_c=$SU(3)$_{L+R}$, the main contribution to $\epsilon_K$, given by the Wilson coefficient $C_4^{sd}$, can be suppressed. However quark masses require the presence of a new scalar field with a vacuum expectation value, whose presence can destabilize the cancellation protecting $C_4$. We have embedded the global symmetry into a larger SU(6) group, showing that a proper spontaneous breaking of the symmetries can occur dynamically. We have made an analysis of the potential, showing which are the representations of the fermions that can trigger this breaking. We have also shown that fermion masses can be reproduced, and the Wilson coefficient $C_4$ can be successfully suppressed, with a compositeness scale $f\sim$TeV. We have also briefly discussed the phenomenology at the LHC of the new light scalar states.

\subsection*{Acknowledgements}

L. D. thanks Kaustubh Agashe for discussions that triggered the realization of this work. We thank Luis Ib\'a\~nez for pointing out references on the physics of axions and Eduardo Andr\'es for useful discussions as well as help with group theoretical issues. We thank Giuliano Panico for questions on magnetic dipole operators. This work was partially supported by ANPCyT PICT 2013-2266.

\appendix
\section{Fundamental representation of SU(6)}\label{Ap-SU(6)}
In this appendix we provide some basic ingredients of the group SU(6).
The generators of SU(6) in the fundamental representation consist of thirty five $6\times 6$ matrices, which are easily described as a combination of three by three blocks, \emph{i.e.}:
\begin{equation}
T^i = \left[\begin{array}{cc} A^i_{3 \times 3} & B^i_{3 \times 3} \\
C^i_{3 \times 3} & D^i_{3 \times 3} 
\end{array}\right] \ ,
\end{equation} 
for the ith generator of $SU(6)$. After the dynamics of the SCFT break $SU(6)$ down, the generators can be organized into those of the preserved subgroup, SU(3)$_L \times$SU(3)$_R\times$U(1) and those of the broken coset.

For the generators of SU(3)$_L \times$SU(3)$_R$, which we will consider to be the first $16$ generators of the basis, $B^i = C^i = 0_{3\times 3}$, $i = 1, ... , 16$. Then, for those of SU(3)$_L$, $A^i =\frac{1}{2} \lambda_i \,$, with the $\lambda$ matrices being the usual Gell-Mann matrices, and $D^i = 0_{3\times 3} \,$, $i = 1,...,8$. On the other hand, for SU(3)$_R$, $A^i =0_{3\times 3}$ and $D^i =\frac{1}{2} \lambda_{i-8}$, $i = 9,...,16$. The generator of U(1) is given by $B^{35} = C^{35} = 0_{3 \times 3}$, $A^{35} = -D^{35} = \frac{1}{2 \sqrt{3}}\, I_{3\times 3}$.

For the coset, the generators arrange into a complex $(3,\overline{3})$. In this case only blocks $B$ and $C$ will be populated by non-zero elements. Furthermore, the $18$ generators of the complex $(3,\overline{3})$ can be organized into two octets and two singlets under the diagonal subgroup of SU(3)$_L \times$SU(3)$_R$, SU(3)$_V$. For the first octet one can choose $A^i = C^i = D^i = 0_{3\times 3}$ and $B^i = K^{i - 16}$, $i = 17,...,24$; for the other $A^i = B^i = D^i = 0_{3\times 3}$ and $C^i = K^{i - 24}$, $i = 25,...,32$. The matrix elements for the $8$ three by three $K^i$ blocks are as follows:
\begin{eqnarray}
& K^1_{(1,3)} = K^2_{(2,3)} = K^3_{(3,1)} = K^4_{(2,1)} = -K^5_{(3,2)} = -K^6_{(1,2)} = \frac{i}{\sqrt{2}},  \nonumber \\
& K^7_{(1,1)} = K^7_{(3,3)} = \frac{i}{2 \sqrt{3}}, \qquad K^7_{(2,2)} = -\frac{i}{\sqrt{3}}, \qquad K^8_{(1,1)} = - K^8_{(3,3)} = - \frac{i}{2}, \nonumber
\end{eqnarray}
with all other elements being zeroes. The two singlets can be described by $A^{33} = D^{33} = A^{34} = D^{34} = 0_{3 \times 3}$, $B^{33} = -C^{33} = i B^{34} = i C^{34} = -\frac{i}{2 \sqrt{3}} \, I_{3 \times 3}$, with $I_{3 \times 3}$ the three by three identity matrix.

\section{Correlators in the symmetric vacuum for the 2-site model}\label{app-correlators}

Let's consider a 2-site model like the one used in this paper or the ones described in \cite{DeCurtis:2011yx,Carena:2014ria,Andres:2015oqa}, where the composite sector has a global symmetry group $G$ that is broken down spontaneously to an $H$ subgroup and fermions transform under full irreducible representations (irreps) of $G$. If we focus solely on the top quark sector of the model, its fermion content consist of two chiral elementary multiplets, one containing the electroweak doublet, $q_L$, and the other containing the electroweak singlet, $u_R$, and two vector-like composite multiplets corresponding to the composite counterparts of the aforementioned elementary multiplets, whose chiral components we call $\Psi_{qL/R}$ and $\Psi_{uL/R}$ respectively. The composite multiplets $\Psi_q$ and $\Psi_u$ are in irreps of $G$, which we shall call $\alpha$ and $\beta$ respectively. Assuming the partial compositeness scheme, we add spurious fields  to the elementary multiplets $q$ or $u$ and embed them in the same irreducible representations of $G$ as the composite ones; these elementary multiplets are called $\psi_q$ and $\psi_u$. Let's define $\alpha_H$ as the set of irreps of $H$ with a non-zero multiplicity in the decomposition of $\alpha$, and similarly define $\beta_H$. For the model to contain Yukawa interactions, it is necessary that $\Gamma := \alpha_H \bigcap \beta_H \neq \emptyset$. Assuming that to be the case, the Lagrangian density for the fermionic sector of this model in the $H$-symmetric vacuum can be described as follows:
\begin{align}
&\mathcal{L} = \mathcal{L}_{\textit{kin}} + \mathcal{L}_{\textit{mass}} + \mathcal{L}_{\textit{mix}} + \mathcal{L}_{\textit{Yuk}} + \mathcal{L}_{\textit{0}} \\
&\mathcal{L}_{\textit{kin}} = \ \overline{\psi}_{qL} \pslash \psi_{qL} +  \ \overline{\psi}_{uR} \pslash \psi_{uR} +  \ \overline{\Psi}_{qL} \pslash \Psi_{qL} + \ \overline{\Psi}_{qR} \pslash \Psi_{qR} +  \ \overline{\Psi}_{uL} \pslash \Psi_{uL} +  \ \overline{\Psi}_{uR} \pslash \Psi_{uR} \nonumber \\
&\mathcal{L}_{\textit{mass}} = -m_{q,1} \ \overline{\Psi}_{qL} \Psi_{qR} - m_{u,1} \ \overline{\Psi}_{uL} \Psi_{uR} + \emph{h.c.}\nonumber\\
&\mathcal{L}_{\textit{mix}} = f_0 \lambda_q \ \overline{\psi}_{qL} \Psi_{qR} + f_0 \lambda_u \ \overline{\psi}_{uR} \Psi_{uL} + \emph{h.c.}\nonumber \\
&\mathcal{L}_{\textit{Yuk}} = f_1 \sum_{r \in \Gamma} \, y_r \, \overline{P_r \, (\Psi_{qL})} P_r \, (\Psi_{uR}) + \emph{h.c.}\nonumber\\
&\mathcal{L}_{0} = - \delta_{\alpha \beta} \, m_y \, \overline{\Psi}_{qR} \Psi_{uL} + \emph{h.c.} \nonumber
\end{align}   
where $P_r$ is the projector from the representation space of $G$ to the subspace associated to the irrep $r$ of $H$. Proceeding in a similar way it is straightforward to include the $d_R$ sector, as well as the light generations of fermions.

We now integrate the composite degrees of freedom using their tree-level equations of motion and calculate the correlators for the elementary fields that transform according to each of the irreps in  $\alpha_H$ or $\beta_H$ (both dynamical and spurious). After integration, the effective fermionic Lagrangian density for the elementary fields in this vacuum takes the form:
\begin{equation}
\mathcal{L}_{\textit{eff}} = \sum_{r \in \alpha_H} \overline{\psi}_{qL}^r \pslash(1 + \Pi_L^r) \psi_{qL}^r + \sum_{s \in \beta_H} \overline{\psi}_{uR}^s \pslash(1 + \Pi_R^s) \psi_{uR}^s + \sum_{t \in \Gamma} (\overline{\psi}_{qL}^t \, \Pi_{LR}^t \, \psi_{uR}^t + \overline{\psi}_{uR}^t \, \Pi_{RL}^t \, \psi_{qL}^t)   
\end{equation} 
with the different correlators given by the following expressions:
\begin{eqnarray}
\Pi_L^r &= - \ f_0^2 \lambda_q^2 \ \frac{p^2 - \, \textbf{I}_{\Gamma}^r (m_{u,1}^2 + \, |f_1 y_r|^2)}{ \delta_{\alpha \beta} [|m_y|^2 |f_1 y_r|^2 - p^2 (|f_1 y_r|^2 + |m_y|^2) + f_1 (y_r^* m_y + y_r m_y^*) m_{q,1} m_{u,1}]  + \left(m_{q,1}^2 - p^2\right) \left(\textbf{I}_{\Gamma}^r m_{u,1}^2 - p^2\right)}\\
\Pi_R^r &= - \ f_0^2 \lambda_u^2 \ \frac{p^2 - \, \textbf{I}_{\Gamma}^r (m_{q,1}^2 + \, |f_1 y_r|^2)}{ \delta_{\alpha \beta} [|m_y|^2 |f_1 y_r|^2 - p^2 (|f_1 y_r|^2 + |m_y|^2) + f_1 (y_r^* m_y + y_r m_y^*) m_{q,1} m_{u,1}]  + \left(\textbf{I}_{\Gamma}^r m_{q,1}^2 - p^2\right) \left(m_{u,1}^2 - p^2\right)}\nonumber\\
\Pi_{LR}^r &= f_0^2 \lambda_q \ \lambda_u \ \frac{\delta_{\alpha \beta} \, m_y (|f_1 y_r|^2 - p^2) + \, \textbf{I}_{\Gamma}^r \, m_{q,1} \, m_{u,1} \, f_1 y_r}{ \delta_{\alpha \beta} [|m_y|^2 |f_1 y_r|^2 - p^2 (|f_1 y_r|^2 + |m_y|^2) + f_1 (y_r^* m_y + y_r m_y^*) m_{q,1} m_{u,1}]  + \left(m_{q,1}^2 - p^2\right) \left(m_{u,1}^2 - p^2\right)}\nonumber\\
\Pi_{RL}^r &= f_0^2 \lambda_q \ \lambda_u \ \frac{\delta_{\alpha \beta} \, m_y^* (|f_1 y_r|^2 - p^2) + \, \textbf{I}_{\Gamma}^r \, m_{q,1} \, m_{u,1} \, f_1 y_r^*}{ \delta_{\alpha \beta} [|m_y|^2 |f_1 y_r|^2 - p^2 (|f_1 y_r|^2 + |m_y|^2) + f_1 (y_r^* m_y + y_r m_y^*) m_{q,1} m_{u,1}]  + \left(m_{q,1}^2 - p^2\right) \left(m_{u,1}^2 - p^2\right)}\nonumber 
\end{eqnarray}
where $\textbf{I}_{\Gamma}^r$ is the characteristic or indicator function of the set $\Gamma$ which takes the value $1$ if $r \in \Gamma$ and is $0$ otherwise.

Note that in the case $\alpha \neq \beta$ and $r \notin \Gamma$ the correlators reduce to:
\begin{equation}
\Pi_L^r = - \frac{\ f_0^2 \lambda_q^2}{ \left(p^2 - m_{q,1}^2\right)} \ ,\qquad\Pi_R^r = - \frac{\ f_0^2 \lambda_u^2}{ \left(p^2 - m_{u,1}^2\right)} \ ,\qquad
\Pi_{LR}^r = \Pi_{RL}^r = 0 \ .
\end{equation}
which are independent of $r$ and thus the same for all irreps of $H$ that don't belong to $\Gamma$. 

Proceeding in a similar way for the gauge fields we obtain the following correlators in the SU(6)-sector:
\begin{align}
\Pi^G_{81}=\Pi^G_{18}=\Pi^G_{11}=\frac{p^2f_{6,0}^2}{2p^2-f_{6,0}^2g_{6,1}^2}\ ,\qquad
\Pi^G_{33}=\frac{f_{6,0}^2(2p^2-f_{6,1}^2g_{6,1}^2)}{2(2p^2-f_{6,0}^2g_{6,1}^2-f_{6,1}^2g_{6,1}^2)}\ .
\end{align}
The case of SO(5)$\times$U(1)$_X$ can be found, for example, in Ref.~\cite{Andres:2015oqa}.

\section{Invariants for the irreps of SO(5)}\label{invariants}

In this Appendix, we present the explicit form of the $F^f_r(s_5)$ and $G^{ff'}_r(s5)$ functions of Eq.~(\ref{eq-corr-f-phys}). Most of these had been calculated previously in the literature, see for example: \cite{Agashe:2004rs,Contino:2006qr,Montull:2013mla,Carena:2014ria,Andres:2015oqa}.

For the singlet of SO(5) we have:

	\begin{center}
		\renewcommand\arraystretch{2.1}
		{\scalebox{0.7}{
		\begin{tabular}{|c||c|c|c|} 
		
\hline
$1_{2/3}$ of SO(5)$\times$U(1)$_X$ &
$u_L$ & $d_L$ &
$u_R$ \\
\hline
$(1,1)_{2/3}$ of SU(2)$_L \times$SU(2)$_R$ &
$F^{u_L}_{11}(s_5) = 0$ & $F^{d_L}_{11}(s_5) = 0$ &
$F^{u_R}_{11}(s_5) = 1$\\
\hline

\end{tabular}}}

	\end{center}

Now, keeping the same layout for the tables but simplifying the notation, the rest of the $F^f_r$ functions are:
\begin{center}
\renewcommand\arraystretch{2.1}
		{\scalebox{0.7}{
\begin{tabular}{|c||c|c|c|} 
\hline
$4_{1/6}$&
$u_L$ & $d_L$ &
$u_R$ \\
\hline
$(2,1)_{1/6}$ &
$\frac{1 + c_5}{2}$ & $\frac{1 + c_5}{2}$ & $\frac{1 - c_5}{2}$\\
\hline
$(1,2)_{1/6}$ &
$\frac{1 - c_5}{2}$ & $\frac{1 - c_5}{2}$ & $\frac{1 + c_5}{2}$\\
\hline
\end{tabular}
\hspace{0.5cm}
\begin{tabular}{|c||c|c|c|} 
\hline
$5_{2/3}$&
$u_L$ & $d_L$ &
$u_R$ \\
\hline
$(2,2)_{2/3}$ &
$\frac{1 + c_5^2}{2}$ & $1$ & $s_5^2$\\
\hline
$(1,1)_{2/3}$ &
$\frac{s_5^2}{2}$ & $0$ & $c_5^2$\\
\hline
\end{tabular}
\hspace{0.5cm}
\begin{tabular}{|c||c|c|c|} 
\hline
$10_{2/3}$&
$u_L$ & $d_L$ &
$u_R$ \\
\hline
$(2,2)_{2/3}$ &
$\frac{1 + c_5^2}{2}$ & $c_5^2$ & $\frac{s_5^2}{2}$\\
\hline
$(3,1)_{2/3}$ &
$\frac{s_5^2}{4}$ & $\frac{s_5^2}{2}$ & $\frac{(c_5^2 - 1)^2}{4}$\\
\hline
$(1,3)_{2/3}$ &
$\frac{s_5^2}{4}$ & $\frac{s_5^2}{2}$ & $\frac{(c_5^2 + 1)^2}{4}$\\
\hline
\end{tabular}}}

\end{center}

\begin{center}
\renewcommand\arraystretch{2.1}
		{\scalebox{0.7}{
\begin{tabular}{|c||c|c|c|} 
\hline
$14_{2/3}$&
$u_L$ & $d_L$ & $u_R$ \\
\hline
$(3,3)_{2/3}$ &
$\frac{1}{4} (2 + 3 c_5^2) s_5^2$ & $s_5^2$ & $\frac{15}{64} (1 - c_5^2 + s_5^2)^2$\\
\hline
$(2,2)_{2/3}$ &
$\frac{1}{2}(c_ 5^4 + s_5^4 + c_5^2 (1 - 2 s_5^2))$ & $c_5^2$ & $\frac{5 c_5^2 s_5^2}{2}$\\
\hline
$(1,1)_{2/3}$ &
$\frac{5 c_5^2 s_5^2}{4}$ & $0$ & $\frac{1}{64} (3 + 5 c_5^2 - 5 s_5^2)^2$\\
\hline
\end{tabular}}}
\end{center}

\begin{center}
\renewcommand\arraystretch{2.1}
		{\scalebox{0.7}{
\begin{tabular}{|c||c|c|c|} 

\hline
$16^{(q_L \in (2,1))}_{1/6}$&
$u_L$ & $d_L$ & $u_R$ \\
\hline
$(3,2)_{1/6}$ &
$-\frac{15}{32} (c_5 - 1)(c_5 + 1)^2$ & $-\frac{15}{32} (c_5 - 1)(c_5 + 1)^2$ & $\frac{15}{32} (c_5 - 1)^2(c_5 + 1)$\\
\hline
$(2,3)_{1/6}$ &
$\frac{15}{32} (c_5 - 1)^2(c_5 + 1)$ & $\frac{15}{32} (c_5 - 1)^2(c_5 + 1)$ & $-\frac{15}{32} (c_5 - 1)(c_5 + 1)^2$\\
\hline
$(2,1)_{1/6}$ &
$\frac{1}{32} (5 c_5 - 1)^2 (c_5 + 1)$ & $\frac{1}{32} (5 c_5 - 1)^2 (c_5 + 1)$ & $-\frac{1}{32} (c_5 - 1) (5 c_5 + 1)^2$\\
\hline
$(1,2)_{1/6}$ &
$-\frac{1}{32} (c_5 - 1) (5 c_5 + 1)^2$ & $-\frac{1}{32} (c_5 - 1) (5 c_5 + 1)^2$ & $\frac{1}{32} (5 c_5 - 1)^2 (c_5 + 1)$\\
\hline
\end{tabular}}}
\end{center}

\begin{center}
\renewcommand\arraystretch{2.1}
		{\scalebox{0.7}{
		\begin{tabular}{|c||c|c|c|} 
		
\hline
$16^{(q_L \in (2,3))}_{1/6}$&
$u_L$ & $d_L$ & $u_R$ \\
\hline
$(3,2)_{1/6}$ &
$\frac{1}{32} (11 - 13 c_5 + 5 c_5^2 - 3 c_5^3)$ & $\frac{1}{32} (11 - 13 c_5 + 5 c_5^2 - 3 c_5^3)$ & $\frac{15}{32} (c_5 - 1)^2 (c_5 + 1)$\\
\hline
$(2,3)_{1/6}$ &
$\frac{1}{32} (11 + 13 c_5 + 5 c_5^2 + 3 c_5^3)$ & $\frac{1}{32} (11 + 13 c_5 + 5 c_5^2 + 3 c_5^3)$ & $-\frac{15}{32} (c_5 - 1) (c_5 + 1)^2$\\
\hline
$(2,1)_{1/6}$ &
$\frac{5}{32} (c_5 - 1)^2 (c_5 + 1)$ & $\frac{5}{32} (c_5 - 1)^2 (c_5 + 1)$ & $-\frac{1}{32} (c_5 - 1) (5 c_5 + 1)^2$\\
\hline
$(1,2)_{1/6}$ &
$-\frac{5}{32} (c_5 - 1) (c_5 + 1)^2$ & $-\frac{5}{32} (c_5 - 1) (c_5 + 1)^2$ & $\frac{1}{32} (5 c_5 - 1)^2 (c_5 + 1)$\\
\hline

\end{tabular}}} 
\end{center}
where we have used the shorthand notation $s_5 = \sin(\tfrac{h_5}{f_5})$ and $c_5 = \cos(\tfrac{h_5}{f_5})$

In a similar fashion, but with a different layout, the expressions for the $G^{u_L u_R}_r$ are shown in the following table:

 \begin{center}
 \renewcommand\arraystretch{2.1}
 		{\scalebox{0.7}{\begin{tabular}{ccc|c|c|c|c|c|c|}
\cline{4-9}
& & & \multicolumn{6}{ c| }{$u_R$ in} \\ \cline{3-9}
& & \multicolumn{1}{ |c|  }{$r \downarrow$} & $1_{2/3}$ & $4_{1/6}$ & $5_{2/3}$ & $10_{2/3}$ & $14_{2/3}$ & $16_{1/6}$ \\ \cline{1-9}
\multicolumn{1}{ |c  }{\multirow{18}{*}{\rotatebox[origin=c]{90}{$u_L$ in}} } & \multicolumn{1}{ |c  }{\multirow{2}{*}{$4_{1/6}$} } &
\multicolumn{1}{ |c| }{$(2,1)_{1/6}$} & - & $\frac{i}{2} s_5$ & - & - & - & $-\frac{1}{8} (5 c_5 + 1) s_5$ \\ \cline{3-9}
\multicolumn{1}{ |c  }{} & \multicolumn{1}{ |c  }{}  &
\multicolumn{1}{ |c| }{$(1,2)_{1/6}$} & - & $-\frac{i}{2} s_5$ & - & - & - & $-\frac{i}{8} (5 c_5 - 1) s_5$ \\ \cline{2-9}
\multicolumn{1}{ |c  }{} & \multicolumn{1}{ |c  }{\multirow{2}{*}{$5_{2/3}$} } &
\multicolumn{1}{ |c| }{$(2,2)_{2/3}$} & - & - & $\frac{s_5 c_5}{\sqrt{2}}$ & $\frac{i s_5}{2}$ & $\frac{i \sqrt{5}}{2} c_5 s_5$ & - \\ \cline{3-9}
\multicolumn{1}{ |c  }{}                      &  \multicolumn{1}{ |c  }{} &
\multicolumn{1}{ |c| }{$(1,1)_{2/3}$} & $-\frac{s_5}{\sqrt{2}}$ & - & $-\frac{s_5 c_5}{\sqrt{2}}$ & - & $\frac{5 s_5^3 - 4 s_5}{4\sqrt{2}}$ & -  \\ \cline{2-9}
\multicolumn{1}{ |c  }{} & \multicolumn{1}{ |c  }{\multirow{3}{*}{$10_{2/3}$} } &
\multicolumn{1}{ |c| }{$(2,2)_{2/3}$} & - & - & $\frac{s_5}{\sqrt{2}}$ & $\frac{i c_5 s_5}{2}$ & $\frac{i \sqrt{5}}{2} c_5^2 s_5$ & -  \\ \cline{3-9}
\multicolumn{1}{ |c  }{}                      &  \multicolumn{1}{ |c  }{} &
\multicolumn{1}{ |c| }{$(3,1)_{2/3}$} & - & - & - & $-\frac{i}{4} s_5 (c_5 - 1)$ & - & -  \\ \cline{3-9}
\multicolumn{1}{ |c  }{}                      &  \multicolumn{1}{ |c  }{} &
\multicolumn{1}{ |c| }{$(1,3)_{2/3}$} & - & - & - & $-\frac{i}{4} s_5 (c_5 + 1)$ & - & -  \\ \cline{2-9}
\multicolumn{1}{ |c  }{} & \multicolumn{1}{ |c  }{\multirow{3}{*}{$14_{2/3}$} } &
\multicolumn{1}{ |c| }{$(3,3)_{2/3}$} & - & - & - & - & $\frac{i 3 \sqrt{5}}{8} c_5 s_5^3 $ & -  \\ \cline{3-9}
\multicolumn{1}{ |c  }{}                      &  \multicolumn{1}{ |c  }{} &
\multicolumn{1}{ |c| }{$(2,2)_{2/3}$} & - & - & $\frac{c_5 s_5}{\sqrt{2}}$ & $\frac{i}{2} (s_5 - 2 s_5^3)$ & $\frac{i \sqrt{5}}{2} c_5 s_5 (1-2 s_5^2)$ & -  \\ \cline{3-9}
\multicolumn{1}{ |c  }{}                      &  \multicolumn{1}{ |c  }{} &
\multicolumn{1}{ |c| }{$(1,1)_{2/3}$} & $-\frac{i \sqrt{5}}{2} c_5 s_5$ & - & $-\frac{i \sqrt{5}}{2} c_5^2 s_5$ & - & $-\frac{i \sqrt{5}}{8} c_5 s_5 (4 - 5 s_5^2)$ & -  \\ \cline{2-9}
\multicolumn{1}{ |c  }{} & \multicolumn{1}{ |c  }{\multirow{4}{*}{$16^{(q_L \in (2,3))}_{1/6}$} } &
\multicolumn{1}{ |c| }{$(3,2)_{1/6}$} & - & - & - & - & - & $-\frac{\sqrt{5}}{32}(c_5-1)(3 c_5-1) s_5$  \\ \cline{3-9}
\multicolumn{1}{ |c  }{}                      &  \multicolumn{1}{ |c  }{} &
\multicolumn{1}{ |c| }{$(2,3)_{1/6}$} & - & - & - & - & - & $\frac{\sqrt{5}}{32}(c_5 + 1)(3 c_5 + 1) s_5$  \\ \cline{3-9}
\multicolumn{1}{ |c  }{}                      &  \multicolumn{1}{ |c  }{} &
\multicolumn{1}{ |c| }{$(2,1)_{1/6}$} & - & $-\frac{i \sqrt{5}}{8}(c_5 - 1) s_5$ & - & - & - & $-\frac{\sqrt{5}}{32}(c_5-1)(5 c_5 + 1) s_5$ \\ \cline{3-9}
\multicolumn{1}{ |c  }{}                      &  \multicolumn{1}{ |c  }{} &
\multicolumn{1}{ |c| }{$(1,2)_{1/6}$} & - & $-\frac{\sqrt{5}}{8}(c_5 + 1) s_5$ & - & - & - & $-\frac{\sqrt{5}}{32}(c_5 + 1)(5 c_5 - 1) s_5$  \\ \cline{2-9}
\multicolumn{1}{ |c  }{} & \multicolumn{1}{ |c  }{\multirow{4}{*}{$16^{(q_L \in (2,1))}_{1/6}$} } &
\multicolumn{1}{ |c| }{$(3,2)_{1/6}$} & - & - & - & - & - & $-\frac{15}{32} s_5^3$  \\ \cline{3-9}
\multicolumn{1}{ |c  }{}                      &  \multicolumn{1}{ |c  }{} &
\multicolumn{1}{ |c| }{$(2,3)_{1/6}$} & - & - & - & - & - & $\frac{15}{32} s_5^3$  \\ \cline{3-9}
\multicolumn{1}{ |c  }{}                      &  \multicolumn{1}{ |c  }{} &
\multicolumn{1}{ |c| }{$(2,1)_{1/6}$} & - & $\frac{i}{8} (5 c_5 - 1) s_5$ & - & - & - & $\frac{1}{32}(1 - 25 c_5^2) s_5$  \\ \cline{3-9}
\multicolumn{1}{ |c  }{}                      &  \multicolumn{1}{ |c  }{} &
\multicolumn{1}{ |c| }{$(1,2)_{1/6}$} & - & $\frac{1}{8} (5 c_5 + 1) s_5$ & - & - & - & $-\frac{1}{32}(1 - 25 c_5^2) s_5$  \\ \cline{1-9}
\end{tabular}}} 
 \end{center}

If a $d_R$ quark was to be introduced in the model, the functions $F^{d_R}_r$ and $G^{d_L d_R}_r$ would also be needed. Their calculation is completely analogous to that of the functions presented here. For the specific case of the $14$-$1$, if we are unwilling to mix the elementary $q_L$ with a second composite multiplet in a different irrep of $SO(5) \times U(1)_X$ then the smallest irrep in which we can embed the $d_R$ is the $10$. For that minimal implementation, we obtain:
\begin{align}
&F^{d_R}_{22}(s_5) = \frac{s_5^2}{2} \ ,\qquad
F^{d_R}_{31}(s_5) =\frac{(1 - c_5)^2}{4}\ , \qquad
F^{d_R}_{13}(s_5) = \frac{(1 + c_5)^2}{4} \ , \nonumber \\
&G^{d_L d_R}_{22}(s_5) = \frac{s_5 c_5}{\sqrt{2}} \ . \nonumber
\end{align}


\begin{thebibliography}{99}
\bibitem{Agashe:2003zs}
  K.~Agashe, A.~Delgado, M.~J.~May and R.~Sundrum,
  JHEP {\bf 0308} (2003) 050
  doi:10.1088/1126-6708/2003/08/050
  [hep-ph/0308036].

\bibitem{Agashe:2006at}
  K.~Agashe, R.~Contino, L.~Da Rold and A.~Pomarol,
  Phys.\ Lett.\ B {\bf 641} (2006) 62
  doi:10.1016/j.physletb.2006.08.005
  [hep-ph/0605341].

\bibitem{Kaplan:1991dc}
  D.~B.~Kaplan,
  Nucl.\ Phys.\ B {\bf 365} (1991) 259.
  doi:10.1016/S0550-3213(05)80021-5

\bibitem{Contino:2004vy}
  R.~Contino and A.~Pomarol,
  JHEP {\bf 0411} (2004) 058
  doi:10.1088/1126-6708/2004/11/058
  [hep-th/0406257].

\bibitem{Gherghetta:2000qt}
  T.~Gherghetta and A.~Pomarol,
  Nucl.\ Phys.\ B {\bf 586} (2000) 141
  doi:10.1016/S0550-3213(00)00392-8
  [hep-ph/0003129].

\bibitem{Agashe:2004cp}
  K.~Agashe, G.~Perez and A.~Soni,
  Phys.\ Rev.\ D {\bf 71} (2005) 016002
  doi:10.1103/PhysRevD.71.016002
  [hep-ph/0408134].

\bibitem{Csaki:2008zd}
  C.~Csaki, A.~Falkowski and A.~Weiler,
  JHEP {\bf 0809} (2008) 008
  doi:10.1088/1126-6708/2008/09/008
  [arXiv:0804.1954 [hep-ph]].

\bibitem{Redi:2011zi}
  M.~Redi and A.~Weiler,
  JHEP {\bf 1111} (2011) 108
  doi:10.1007/JHEP11(2011)108
  [arXiv:1106.6357 [hep-ph]].

\bibitem{Panico:2015jxa}
  G.~Panico and A.~Wulzer,
  Lect.\ Notes Phys.\  {\bf 913} (2016) pp.1
  doi:10.1007/978-3-319-22617-0
  [arXiv:1506.01961 [hep-ph]].

\bibitem{Delaunay:2010dw}
  C.~Delaunay, O.~Gedalia, S.~J.~Lee, G.~Perez and E.~Pont\'on,
  Phys.\ Rev.\ D {\bf 83} (2011) 115003
  doi:10.1103/PhysRevD.83.115003
  [arXiv:1007.0243 [hep-ph]].

\bibitem{Panico:2016ull}
  G.~Panico and A.~Pomarol,
  JHEP {\bf 1607} (2016) 097
  doi:10.1007/JHEP07(2016)097
  [arXiv:1603.06609 [hep-ph]].

\bibitem{Bauer:2011ah}
  M.~Bauer, R.~Malm and M.~Neubert,
  Phys.\ Rev.\ Lett.\  {\bf 108} (2012) 081603
  doi:10.1103/PhysRevLett.108.081603
  [arXiv:1110.0471 [hep-ph]].

\bibitem{DaRold:2012sz} 
  L.~Da Rold, C.~Delaunay, C.~Grojean and G.~Perez,
  JHEP {\bf 1302}, 149 (2013)
  doi:10.1007/JHEP02(2013)149
  [arXiv:1208.1499 [hep-ph]].

\bibitem{Agashe:2004rs}
  K.~Agashe, R.~Contino and A.~Pomarol,
  Nucl.\ Phys.\ B {\bf 719} (2005) 165
  [hep-ph/0412089].

\bibitem{Contino:2006nn}
  R.~Contino, T.~Kramer, M.~Son and R.~Sundrum,
  JHEP {\bf 0705} (2007) 074
  doi:10.1088/1126-6708/2007/05/074
  [hep-ph/0612180].

\bibitem{Foadi:2010bu}
  R.~Foadi, J.~T.~Laverty, C.~R.~Schmidt and J.~H.~Yu,
  JHEP {\bf 1006} (2010) 026
  doi:10.1007/JHEP06(2010)026
  [arXiv:1001.0584 [hep-ph]].

\bibitem{Panico:2011pw}
  G.~Panico and A.~Wulzer,
  JHEP {\bf 1109} (2011) 135
  doi:10.1007/JHEP09(2011)135
  [arXiv:1106.2719 [hep-ph]].

\bibitem{DeCurtis:2011yx}
  S.~De Curtis, M.~Redi and A.~Tesi,
  JHEP {\bf 1204} (2012) 042
  doi:10.1007/JHEP04(2012)042
  [arXiv:1110.1613 [hep-ph]].

\bibitem{Andres:2015oqa}
  E.~C.~Andr\'es, L.~Da Rold and I.~A.~Davidovich,
  JHEP {\bf 1603} (2016) 152
  doi:10.1007/JHEP03(2016)152
  [arXiv:1509.04726 [hep-ph]].

\bibitem{Contino:2006qr}
  R.~Contino, L.~Da Rold and A.~Pomarol,
  Phys.\ Rev.\ D {\bf 75} (2007) 055014
  doi:10.1103/PhysRevD.75.055014
  [hep-ph/0612048].

\bibitem{Montull:2013mla}
  M.~Montull, F.~Riva, E.~Salvioni and R.~Torre,
  Phys.\ Rev.\ D {\bf 88} (2013) 095006
  doi:10.1103/PhysRevD.88.095006
  [arXiv:1308.0559 [hep-ph]].

\bibitem{Carena:2014ria}
  M.~Carena, L.~Da Rold and E.~Pont\'on,
  JHEP {\bf 1406} (2014) 159
  doi:10.1007/JHEP06(2014)159
  [arXiv:1402.2987 [hep-ph]].

\bibitem{Grossman:1999ra}
  Y.~Grossman and M.~Neubert,
  Phys.\ Lett.\ B {\bf 474} (2000) 361
  doi:10.1016/S0370-2693(00)00054-X
  [hep-ph/9912408].

\bibitem{Luty:2004ye}
  M.~A.~Luty and T.~Okui,
  JHEP {\bf 0609} (2006) 070
  doi:10.1088/1126-6708/2006/09/070
  [hep-ph/0409274].

\bibitem{Rattazzi:2008pe}
  R.~Rattazzi, V.~S.~Rychkov, E.~Tonni and A.~Vichi,
  JHEP {\bf 0812} (2008) 031
  doi:10.1088/1126-6708/2008/12/031
  [arXiv:0807.0004 [hep-th]].

\bibitem{Panico:2012uw}
  G.~Panico, M.~Redi, A.~Tesi and A.~Wulzer,
  JHEP {\bf 1303} (2013) 051
  [arXiv:1210.7114 [hep-ph]].

\bibitem{Agashe:2009di}
  K.~Agashe and R.~Contino,
  Phys.\ Rev.\ D {\bf 80} (2009) 075016
  [arXiv:0906.1542 [hep-ph]].

\bibitem{Mrazek:2011iu}
  J.~Mrazek, A.~Pomarol, R.~Rattazzi, M.~Redi, J.~Serra and A.~Wulzer,
  Nucl.\ Phys.\ B {\bf 853} (2011) 1
  [arXiv:1105.5403 [hep-ph]].

\bibitem{Son:2003et}
  D.~T.~Son and M.~A.~Stephanov,
  Phys.\ Rev.\ D {\bf 69} (2004) 065020
  doi:10.1103/PhysRevD.69.065020
  [hep-ph/0304182].

\bibitem{Alvarez:2016ljl}
  E.~Alvarez, L.~Da Rold, J.~Mazzitelli and A.~Szynkman,
  arXiv:1610.08451 [hep-ph].

\bibitem{Agashe:2008uz}
  K.~Agashe, A.~Azatov and L.~Zhu,
  Phys.\ Rev.\ D {\bf 79} (2009) 056006
  doi:10.1103/PhysRevD.79.056006
  [arXiv:0810.1016 [hep-ph]].

\bibitem{Azatov:2011qy}
  A.~Azatov and J.~Galloway,
  Phys.\ Rev.\ D {\bf 85} (2012) 055013
  doi:10.1103/PhysRevD.85.055013
  [arXiv:1110.5646 [hep-ph]].

\bibitem{Delaunay:2012cz}
  C.~Delaunay, J.~F.~Kamenik, G.~Perez and L.~Randall,
  JHEP {\bf 1301} (2013) 027
  doi:10.1007/JHEP01(2013)027
  [arXiv:1207.0474 [hep-ph]].

\bibitem{Konig:2014iqa}
  M.~König, M.~Neubert and D.~M.~Straub,
  Eur.\ Phys.\ J.\ C {\bf 74} (2014) 2945
  [arXiv:1403.2756 [hep-ph]].

\bibitem{Aad:2014ioa}
  G.~Aad {\it et al.} [ATLAS Collaboration],
  Phys.\ Rev.\ Lett.\  {\bf 113} (2014) no.17,  171801
  doi:10.1103/PhysRevLett.113.171801
  [arXiv:1407.6583 [hep-ex]].

\bibitem{CMS-PAS-HIG-14-037}
CMS-PAS-HIG-14-037, CMS Collaboration.

\bibitem{Ibanez:2015uok}
  L.~E.~Ibanez and V.~Martin-Lozano,
  JHEP {\bf 1607} (2016) 021
  doi:10.1007/JHEP07(2016)021
  [arXiv:1512.08777 [hep-ph]].

\bibitem{Herraez:2016dxn}
  A.~Herraez and L.~E.~Ibanez,
  JHEP {\bf 1702} (2017) 109
  doi:10.1007/JHEP02(2017)109
  [arXiv:1610.08836 [hep-th]].

\bibitem{Aad:2014aqa}
  G.~Aad {\it et al.} [ATLAS Collaboration],
  Phys.\ Rev.\ D {\bf 91} (2015) no.5,  052007
  doi:10.1103/PhysRevD.91.052007
  [arXiv:1407.1376 [hep-ex]].

\bibitem{Khachatryan:2015sja}
  V.~Khachatryan {\it et al.} [CMS Collaboration],
  Phys.\ Rev.\ D {\bf 91} (2015) no.5,  052009
  doi:10.1103/PhysRevD.91.052009
  [arXiv:1501.04198 [hep-ex]].

\bibitem{Sirunyan:2016iap}
  A.~M.~Sirunyan {\it et al.} [CMS Collaboration],
  Phys.\ Lett.\ B
  doi:10.1016/j.physletb.2017.02.012
  [arXiv:1611.03568 [hep-ex]].

\bibitem{Barbieri:1987fn}
  R.~Barbieri and G.~F.~Giudice,
  Nucl.\ Phys.\ B {\bf 306} (1988) 63.

\bibitem{Anderson:1994dz} 
  G.~W.~Anderson and D.~J.~Castano,
  Phys.\ Lett.\ B {\bf 347}, 300 (1995)
  [hep-ph/9409419].
  
\end{thebibliography}
\end{document}